\pdfoutput=1

\documentclass[%
reprint,
superscriptaddress,
amsmath,amssymb,
aps,10pt,twocolumn,prb]{revtex4-2}

\usepackage{graphicx}%
\usepackage{dcolumn}%
\usepackage{bm}%

\usepackage[normalem]{ulem}
\usepackage[svgnames,psnames]{xcolor} %
\usepackage[colorlinks,citecolor=DarkGreen,linkcolor=FireBrick,linktocpage,unicode]{hyperref} %
\usepackage{textcomp}  
\usepackage{rotating} 
\usepackage{comment}

\begin{document}

\preprint{APS/123-QED}

\title{Predictive power of polynomial machine learning potentials for liquid states in 22 elemental systems}

\author{Hayato \surname{Wakai}}
\email{wakai@cms.mtl.kyoto-u.ac.jp}
\author{Atsuto \surname{Seko}}
\email{seko@cms.mtl.kyoto-u.ac.jp}
\author{Hirosato \surname{Izuta}}
\author{Takayuki \surname{Nishiyama}}
\author{Isao \surname{Tanaka} \\ \small{\textit{Department of Materials Science and Engineering, Kyoto University, Kyoto 606-8501, Japan}}}
\date{\today}

\begin{abstract}
The polynomial machine learning potentials (MLPs) described by polynomial rotational invariants have been systematically developed for various systems and used in diverse applications in crystalline states. In this study, we systematically investigate the predictive power of the polynomial MLPs for liquid structural properties in 22 elemental systems with diverse chemical bonding properties, including those showing anomalous melting behavior, such as Si, Ge, and Bi. We compare liquid structural properties obtained from molecular dynamics simulations using the density functional theory (DFT) calculation, the polynomial MLPs, and other interatomic potentials in the literature. The current results demonstrate that the polynomial MLPs consistently exhibit high predictive power for liquid structural properties with the same accuracy as that of typical DFT calculations.

\end{abstract}

\maketitle

\section{Introduction}

Molecular dynamics (MD) simulations are powerful tools for performing atomistic simulations in a wide range of applications. For MD simulations in liquid and liquid-like disordered states, empirical interatomic potentials such as pair potentials (e.g., Lennard--Jones (LJ) potential \cite{LJ_potential}), embedded atom method (EAM) potentials \cite{PhysRevB.29.6443}, and Stillinger--Weber (SW) potentials \cite{PhysRevB.31.5262} have been commonly employed \cite{doi:10.1080/14786430310001613264,doi:10.1080/14786430802206482, PhysRevB.35.9120, Sastry2003}. Although the empirical interatomic potentials allow us to extend the simulation time and the number of atoms in the simulation cell significantly, their simplistic models often restrict their application and target system, as will also be demonstrated in this study. \textit{Ab initio} MD (AIMD) can be an alternative way to perform accurate atomistic simulations \cite{PhysRevB.47.558, marx2009ab}. However, its high computational cost limits its application.

Machine learning potentials (MLPs)
\cite{
Lorenz2004210,
behler2007generalized,
bartok2010gaussian,
behler2011atom,
han2017deep,
258c531ae5de4f5699e2eec2de51c84f,
PhysRevB.96.014112,
PhysRevB.90.104108,
PhysRevX.8.041048,
PhysRevLett.114.096405,
PhysRevB.95.214302,
PhysRevB.90.024101,
PhysRevB.92.054113,
PhysRevMaterials.1.063801,
Thompson2015316,
wood2018extending,
PhysRevMaterials.1.043603,
doi-10.1137-15M1054183,
doi:10.1063/1.5126336,
Freitas2022} 
have been in increased demand for accurate and efficient large-scale simulations that are prohibitively expensive using the density functional theory (DFT) calculation. The MLPs represent interatomic interactions with systematic structural features and flexible machine learning models such as artificial neural networks, Gaussian process models, and linear models. Because the MLPs are typically developed from extensive DFT datasets, they should have high predictive power for structures close to those in the datasets. They have been applied to efficient and accurate calculations for liquid states \cite{PhysRevX.8.041048, PhysRevB.95.094203,doi:10.1073/pnas.1815117116}.

The polynomial MLP is an approach to develop MLPs that are accurate for a wide variety of structures \cite{PhysRevB.99.214108, PhysRevB.102.174104,doi:10.1063/5.0129045}. The polynomial MLP is described as polynomial rotational invariants systematically derived from order parameters in terms of radial and spherical harmonic functions. Because simple polynomial functions are employed instead of artificial neural networks and Gaussian process models, the descriptive power for the potential energy is strongly dependent on their polynomial forms. At the same time, efficient model estimations can be achieved using linear regressions supported by powerful libraries for linear algebra \cite{doi:10.1063/5.0129045}. Moreover, the force and stress tensor components in DFT training datasets can be considered in a straightforward manner \cite{doi:10.1063/5.0129045}. Even when considering the force and stress tensor components as training data entries, it is possible to estimate model coefficients efficiently using fast linear regressions.

Following these advantages, polynomial MLPs have been systematically developed for various systems. Various developed polynomial MLPs with different trade-offs between accuracy and computational efficiency are available in the Polynomial Machine Learning Potential Repository \cite{MachineLearningPotentialRepository}. Currently, the repository contains MLPs for 48 elemental and 120 binary alloy systems. They have been used for applications in crystalline states \cite{PhysRevB.99.214108, PhysRevB.102.174104, PhysRevMaterials.4.123607, FUJII2022111137}, which indicates that they can enable us to predict properties accurately for various crystal structures. In this study, we systematically investigate the predictive power of polynomial MLPs for liquid states in 22 elemental systems, i.e., Li, Be, Na, Mg, Al, Si, Ti, V, Cr, Cu, Zn, Ga, Ge, Ag, Cd, In, Sn, Au, Hg, Tl, Pb, and Bi. We compare the structural quantities commonly used to describe liquid structures computed using the DFT calculations, the polynomial MLPs, and other interatomic potentials. The current targets include elemental systems known to exhibit anomalous melting behavior, such as Si, Ge, Sn, Ga, and Bi \cite{Statistical_Physics_of_Crystals_and_Liquids, anomalous_liquid, anomalous_liquid2}. As will be shown below, complex descriptions for the potential energy using the MLPs are essential for accurately predicting liquid structural quantities.

\section{Methodology}
\subsection{Polynomial machine learning potentials}

In this section, we present the formulation of the polynomial MLP in elemental systems, which can be simplified from the formulation for multicomponent systems \cite{PhysRevB.102.174104,doi:10.1063/5.0129045}.
The short-range part of the potential energy for a structure, $E$, is assumed to be decomposed as $E = \sum_i E^{(i)}$, where $E^{(i)}$ denotes the contribution of interactions between atom $i$ and its neighboring atoms within a given cutoff radius $r_c$, referred to as the atomic energy.
The atomic energy is then approximately given by a function of invariants $\{d_{m}^{(i)}\}$ with any rotations centered at the position of atom $i$ as
\begin{equation}
\label{sscha:Eqn-atomic-energy-features}
E^{(i)} = F \left( d_1^{(i)}, d_2^{(i)}, \cdots \right),
\end{equation}
where $d_{m}^{(i)}$ can be referred to as a structural feature for modeling the potential energy.
The polynomial MLP adopts polynomial invariants of the order parameters representing the neighboring atomic density as structural features and employs polynomial functions as function $F$.

When the neighboring atomic density is described by radial functions $\{f_n\}$ and spherical harmonics $\{Y_{lm}\}$, a $p$th-order polynomial invariant for radial index $n$ and set of angular numbers $\{l_1,l_2,\cdots,l_p\}$ is given by a linear combination of products of $p$ order parameters, expressed as
\begin{widetext}
\begin{equation}
\label{sscha:Eqn-invariant-form}
d_{nl_1l_2\cdots l_p,(\sigma)}^{(i)} =
\sum_{m_1,m_2,\cdots, m_p} c^{l_1l_2\cdots l_p,(\sigma)}_{m_1m_2\cdots m_p}
a_{nl_1m_1}^{(i)} a_{nl_2m_2}^{(i)} \cdots a_{nl_pm_p}^{(i)},
\end{equation}
\end{widetext}
where the order parameter $a^{(i)}_{nlm}$ is component $nlm$ of the neighboring atomic density of atom $i$.
The coefficient set $\{c^{l_1l_2\cdots l_p,(\sigma)}_{m_1m_2\cdots m_p}\}$ ensures that the linear combinations are invariant for arbitrary rotations, which can be enumerated using group theoretical approaches such as the projection operator method \cite{el-batanouny_wooten_2008, PhysRevB.99.214108}.
In terms of fourth- and higher-order polynomial invariants, multiple linear combinations are linearly independent for most of the set $\{l_1,l_2,\cdots,l_p\}$.
They are distinguished by index $\sigma$ if necessary.

Here, the radial functions are Gaussian-type ones expressed by
\begin{equation}
f_{n}(r)=\exp\left[-\beta_n(r-r_n)^{2}\right] f_c(r),
\end{equation}
where $\beta_n$ and $r_n$ denote given parameters.
The cutoff function $f_c$ ensures the smooth decay of the radial function. 
The current MLP employs a cosine-based cutoff function expressed as
\begin{eqnarray}
f_c(r) = \left\{
\begin{aligned}
& \frac{1}{2} \left[ \cos \left( \pi \frac{r}{r_c} \right) + 1\right] & (r \le r_c)\\
& 0 & (r > r_c)
\end{aligned}
\right ..
\end{eqnarray}
The order parameter of atom $i$, $a_{nlm}^{(i)}$, is approximately evaluated from the neighboring atomic distribution of atom $i$ as
\begin{equation}
a_{nlm}^{(i)} = \sum_{\{j | r_{ij} \leq r_c\} }
f_n(r_{ij}) Y_{lm}^* (\theta_{ij}, \phi_{ij}),
\end{equation}
where $(r_{ij}, \theta_{ij}, \phi_{ij})$ denotes the spherical coordinates of neighboring atom $j$ centered at the position of atom $i$.
Note that this approximation for the order parameters ignores the nonorthonormality of the Gaussian-type radial functions, but it is acceptable in developing the polynomial MLP \cite{PhysRevB.99.214108}.

Given a set of structural features $D^{(i)} = \{d_1^{(i)},d_2^{(i)},\cdots\}$, the polynomial function $F_\xi$ composed of all combinations of $\xi$ structural features is represented as
\begin{eqnarray}
F_1 \left(D^{(i)}\right) &=& \sum_{s} w_{s} d_{s}^{(i)}, \nonumber \\
F_2 \left(D^{(i)}\right) &=& \sum_{\{st\}} w_{st} d_{s}^{(i)} d_{t}^{(i)}, \\
F_3 \left(D^{(i)}\right) &=& \sum_{\{stu\}} w_{stu} d_{s}^{(i)} d_{t}^{(i)} d_{u}^{(i)} \nonumber,
\end{eqnarray}
where $w$ denotes a regression coefficient.
A polynomial of the polynomial invariants $D^{(i)}$ is then described as
\begin{equation}
\label{Eqn-polynomial-model1}
E^{(i)} = F_1 \left( D^{(i)} \right) + F_2 \left( D^{(i)} \right)
+ F_3 \left( D^{(i)} \right) + \cdots.
\end{equation}
The current models have no constant terms, which means that the atomic energy is measured from the energies of isolated atoms.
In addition to the model given by Eq. (\ref{Eqn-polynomial-model1}), simpler models composed of a linear polynomial of structural features and a polynomial of a subset of the structural features are also introduced, such as
\begin{eqnarray}
\label{Eqn-polynomial-model2}
%E^{(i)} &=& F_1\left(D_{\rm pair}^{(i)} \right)
%+ F_2\left(D_{\rm pair}^{(i)} \right)
%+ F_3\left(D_{\rm pair}^{(i)} \right) \nonumber \\
%E^{(i)} &=& F_1 \left( D^{(i)} \right) + F_2 \left( D_{\rm pair}^{(i)} \right)
%\\
E^{(i)} &=& F_1 \left( D^{(i)} \right)
+ F_2 \left( D_{\rm pair}^{(i)} \cup D_2^{(i)} \right),
\end{eqnarray}
where subsets of $D^{(i)}$ are denoted by
\begin{eqnarray}
D_{\rm pair}^{(i)} = \{d_{n0}^{(i)}\}, D_2^{(i)} = \{d_{nll}^{(i)}\}.
\end{eqnarray}

Note that the polynomial MLP is equivalent to a spectral neighbor analysis potential (SNAP) \cite{Thompson2015316} when the linear polynomial model with up to third-order invariants is expressed as
\begin{equation}
E^{(i)} = F_1 \left( D_{\rm pair}^{(i)} \cup D_2^{(i)} \cup D_3^{(i)} \right),
\end{equation}
where subset $D_3^{(i)}$ is given as $D_3^{(i)} = \{d_{nl_1l_2l_3}^{(i)}\}$.
Similarly, the current formulation includes a quadratic SNAP \cite{wood2018extending}, which is an extension of the SNAP.
In addition, linear polynomial models using polynomial invariants are analogous to the formulation of the atomic cluster expansion  \cite{PhysRevB.99.014104}.

Each of the polynomial MLPs was created from a training dataset using \textsc{pypolymlp} \cite{doi:10.1063/5.0129045, PYPOLYMLP}, and its prediction errors for the energy, force components, and stress tensors were estimated using a test dataset. The training and test datasets for each elemental metal were generated as follows. 
First, we fully optimized the atomic positions and lattice constants of 86 prototype structures \cite{PhysRevB.99.214108} using the DFT calculation. 
They comprise single elements with the zero oxidation state from the Inorganic Crystal Structure Database (ICSD) \cite{bergerhoff1987crystal}, including metallic closed-packed structures, covalent structures, layered structures, and structures reported as high-pressure phases. 
Then, 13000--15000 structures were generated from the optimized prototype structures, and they were randomly divided into training and test datasets at a ratio of nine to one. 
Each structure was constructed by randomly introducing lattice expansions, lattice distortions, and atomic displacements into a supercell of an optimized prototype structure. 
No structural data in liquid states, such as structural trajectories in MD simulations at high temperatures, were used to develop the polynomial MLPs.

Non-spin-polarized DFT calculations were performed for structures in the datasets using the plane-wave-basis projector augmented wave (PAW) method \cite{PhysRevB.50.17953,PhysRevB.59.1758} within the Perdew--Burke--Ernzerhof (PBE) exchange-correlation functional \cite{PhysRevLett.77.3865}, as implemented in the \textsc{vasp} code \cite{PhysRevB.47.558, PhysRevB.54.11169,KRESSE199615}.
The cutoff energy was set to 300 eV.
The total energies converged to less than 10$^{-3}$ meV/supercell.
The allowed spacing between $k$-points was approximately set to 0.09 $\rm{\AA}^{-1}$.
The atomic positions and lattice constants of the prototype structures were optimized until the residual forces were less than 10$^{-2}$ eV/\AA. 
The PAW potentials used in DFT calculations are listed in Appendix \ref{appendixA}, and these PAW potentials include scalar-relativistic corrections. 
Spin-orbit coupling was not considered in all the elemental systems.

Regression coefficients of potential energy models were estimated using linear ridge regression.
The energy values and force components in the training dataset were used as observations in the regression.
The ridge regularization parameter was optimized to minimize the prediction error for the test dataset.

The accuracy and computational efficiency of the polynomial MLP greatly depend on the input parameters, such as the cutoff radius and the number of order parameters. Therefore, a systematic grid search was conducted to find their optimal values in each system. As indicated in Ref. \cite{PhysRevB.99.214108}, the accuracy and computational efficiency are conflicting properties, hence, a set of Pareto-optimal MLPs with different trade-offs between the accuracy and computational efficiency was obtained from the grid search.

\begin{figure*}[!tb]
    \centering
    \includegraphics[width=0.98\linewidth]{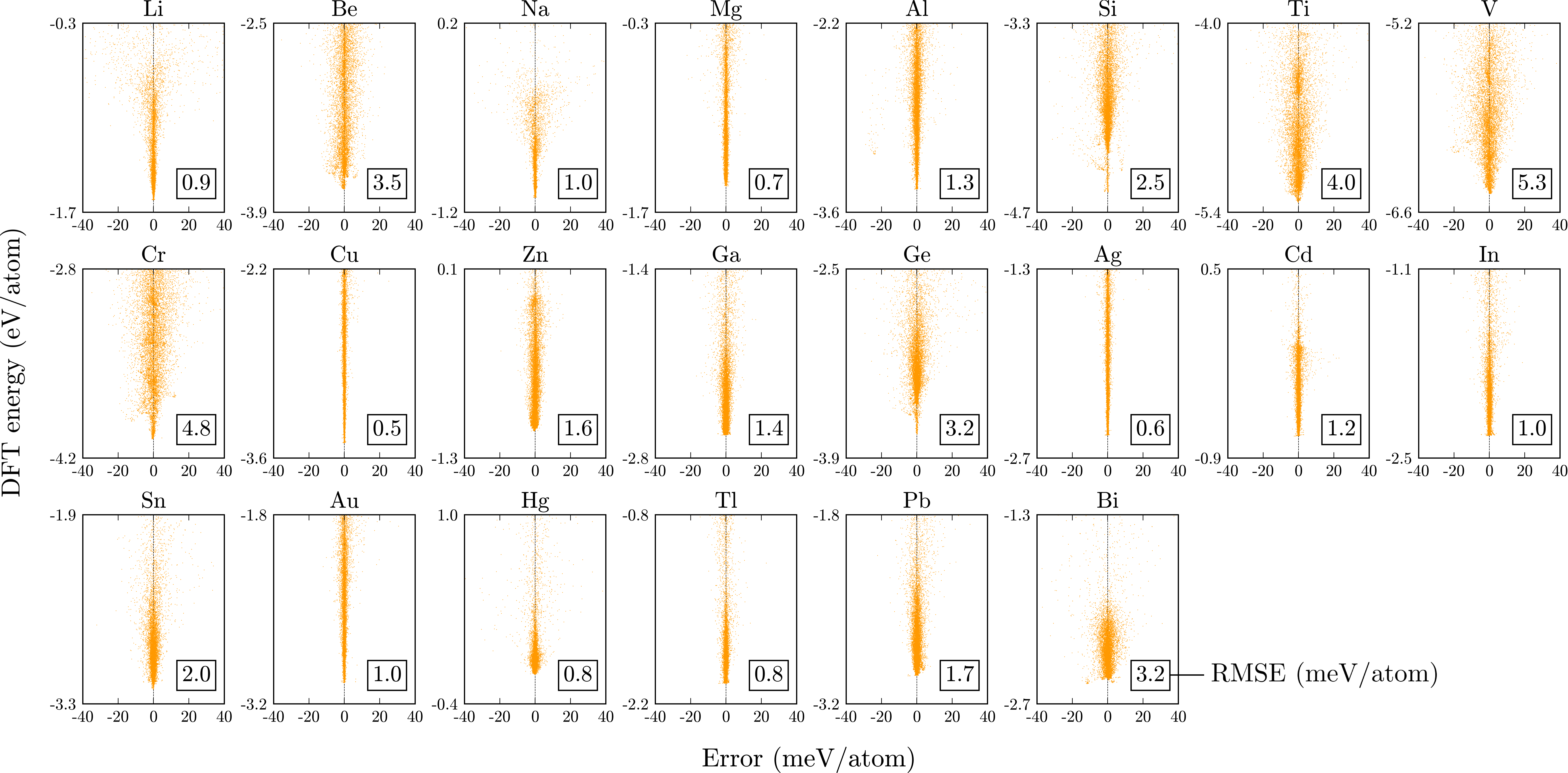}
    \caption{Distribution of the energies of structures in the training and test datasets computed using the DFT calculation and those calculated using the polynomial MLP, which has the lowest prediction error. The vertical axis range is fixed to 1.4 eV/atom, although structures with energy values higher than the energy range are included in the datasets. The numerical values enclosed in the squares represent the root square mean errors (RMSEs) for the energy, which are estimated using the test datasets.}
    \label{fig:sum_e_dist}
\end{figure*}

\begin{figure*}[!p]
    \centering
    \includegraphics[width=0.90\linewidth]{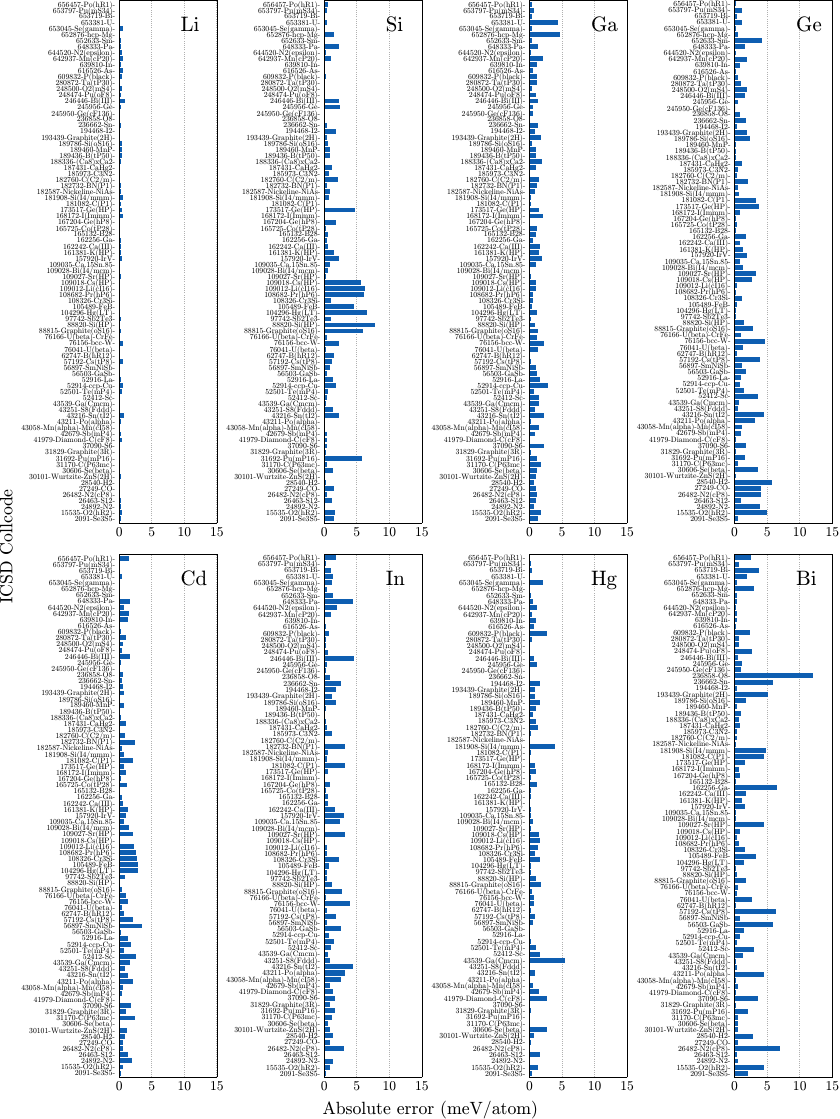}
    \caption{Absolute prediction errors of the cohesive energy for 86 prototype structures in elemental Li, Si, Ga, Ge, Cd, In, Hg, and Bi. The absolute prediction errors for the other systems can be found in the supplemental material.
        }
    \label{fig:sum_proto}
\end{figure*}

Figure \ref{fig:sum_e_dist} shows the distribution of the energies of structures in the training and test datasets computed using the DFT calculation and those calculated using the polynomial MLP, which has the lowest prediction error. 
The polynomial MLP reveals a narrow distribution of errors. 
Figure \ref{fig:sum_proto} shows the absolute prediction errors of the cohesive energy for various prototype structures. 
The polynomial MLP exhibits minor errors for almost all prototype structures. 
These results indicate that the polynomial MLP is accurate for many typical structures and their derivatives containing diverse neighborhood environments and coordination numbers.

\subsection{Computational procedures for MD simulations}\label{2B}
Multiple Pareto-optimal MLPs developed using the above procedure are available in the repository \cite{doi:10.1063/5.0129045, MachineLearningPotentialRepository}. They show different trade-offs between accuracy and computational efficiency. 
Although a polynomial MLP is generally chosen from the set of Pareto-optimal MLPs for performing atomistic simulations, we examine the accuracy of all Pareto-optimal MLPs for liquid states. 

We perform MD simulations using the polynomial MLPs and the DFT calculation for elemental Li, Be, Na, Mg, Al, Si, Ti, V, Cr, Cu, Zn, Ga, Ge, Ag, Cd, In, Sn, Au, Hg, Tl, Pb, and Bi at temperatures close to and above their melting temperatures. We also employ empirical interatomic potentials and other MLPs available in open repositories such as OpenKim \cite{OpenKim} and the interatomic potential repository \cite{NIST}. 

The MD simulations were performed within the NVT ensemble, employing the Nose--Hoover thermostat \cite{10.1063/1.447334, PhysRevA.31.1695} to control the temperature. MD simulations were carried out using the LAMMPS code \cite{LAMMPS}. The current computational procedure for performing a single MD simulation is as follows. We first generate a cubic periodic cell with 125 atoms arranged on a 5 × 5 × 5 regular grid. The cell volume is given such that the cell density corresponds to the experimental density at its melting temperature, as reported in Ref. \cite{liquid_dence}. The cell density for the elemental Hg is exceptionally given as 12.4 $\rm{g/cm^3}$. This cell density value was employed by Kresse and Hafner \cite{PhysRevB.55.7539}. The atomic configurations are then equilibrated using MD run for 3 ps at a temperature typically 700 K higher than the experimental melting temperature to obtain a snapshot in the liquid state. The MD time step is set to 3 fs. The atomic configurations are further equilibrated from the snapshot structure using MD run for at least 3 ps at the target temperature. Finally, we perform MD run for 15 ps at the target temperature and calculate structural quantities as ensemble averages over the MD trajectory.

For the AIMD simulations, non-spin-polarized DFT calculations were performed using the plane-wave-basis PAW method \cite{PhysRevB.50.17953,PhysRevB.59.1758} within the PBE exchange-correlation functional \cite{PhysRevLett.77.3865}, as implemented in the \textsc{vasp} code \cite{PhysRevB.47.558,PhysRevB.54.11169,KRESSE199615}. 
The cutoff energy was set to 400 eV. 
The integration in the reciprocal space was performed at the $\Gamma$-point only. 
The PAW potentials listed in Appendix \ref{appendixA} were utilized to perform AIMD simulations, and these PAW potentials include scalar-relativistic corrections. 
These PAW potentials were the same as those used for constructing the polynomial MLPs. 
The DFT calculations were performed without considering spin-orbit coupling.

\subsection{Structural quantities for liquid states}

\subsubsection{Radial distribution function}

The radial distribution function (RDF), denoted as $g(r)$, has been widely used for describing liquid structures quantitatively \cite{Allen, tuckerman2023statistical}. The RDF characterizes the spatial distribution of atoms as a function of distance $r$ from a reference atom relative to the probability for a completely random distribution. In practice, the RDF is approximately calculated as a histogram with a given bin width. Here, we use the bin width of 0.1 $\rm{\AA}$ to evaluate the RDF.

\subsubsection{Bond-angle distribution function}

The bond-angle distribution function (BADF), denoted as $g(\theta)$, has been employed to analyze the local orientational order in liquid and disordered states \cite{PhysRevLett.63.2240, 10.1063/1.2431173, MUNETOH2007334}. The BADF can be defined as the probability distribution of bond angles formed by two neighboring atoms within a given cutoff distance. In this study, the cutoff distance is given as 1.4 times the distance corresponding to the first peak in the RDF. Moreover, the BADF is also practically evaluated as a histogram using the bin width of one degree.

\subsubsection{Running coordination number}

The coordination number is often given as the integration of the RDF up to the distance corresponding to the first minimum in the
RDF \cite{PhysRevB.47.558, 10.1063/1.2431173, Statistical_Physics_of_Crystals_and_Liquids}. However, it is problematic to determine the distance of the first minimum precisely in some systems exhibiting flat first minima. In such a case, the coordination number cannot be estimated robustly because of the uncertainty of the first minimum position. Therefore, we employ the running coordination number (CN) \cite{tuckerman2023statistical} as a structural quantity, which is defined as 
\begin{align}
N_{\rm{cn}}(r) =\frac{ 4 \pi N}{V} \int_{0}^{r} \tilde{r}^2 g(\tilde{r})  {\rm{d}}\tilde{r},
\end{align}
where $N$ is the number of atoms in the system, and $V$ is the system volume. The running CN provides the average number of atoms coordinating a given atom out to a distance $r$.

\subsubsection{Bond-orientational order parameters}

The bond-orientational order parameters (BOOPs) proposed by Steinhardt \textit{et al.} \cite{BOO_param} have been used for characterizing the local orientational order in liquid and disordered states \cite{10.1063/1.2977970, 10.1063/1.4955305, WINCZEWSKI201957}. The BOOPs are equivalent to second-order polynomial invariants of spherical harmonics with any rotation. Therefore, the definition of the BOOPs is similar to the second-order polynomial invariants used in the polynomial MLPs. The order parameter around the central atom $i$, $Q_l(i)$, is given by second-order polynomial invariants of $Q_{lm}(i)$ as 
\begin{align}
Q_l(i)=\sqrt{\frac{4\pi}{2l+1}\sum\limits_{m=-l}^l{\mid Q_{lm}(i) \mid}^2 },
\end{align}
where $Q_{lm}(i)$ denotes the spherical harmonic functions $Y_{lm}$ averaged over its neighboring atoms described by 
\begin{align}
Q_{lm}(i) = \frac{1}{N_{\rm{neigh}}}\sum\limits_{j\in{\rm{neighbor}}}Y_{lm}(\theta_{ij},\varphi_{ij}).
\end{align}
Angles $\theta_{ij}$ and $\phi_{ij}$ give the azimuthal and polar angles of the spherical coordinates of neighboring atom $j$ centered at the position of atom $i$. The BOOP of angular number $l$ is then defined as the average of $Q_l(i)$ over all atoms, expressed as 
\begin{align}
Q_{l} = \frac{1}{N}\sum\limits_{i=1}^N Q_{l}(i).
\end{align}
We employ the ensemble average of the BOOP, $\left\langle Q_l \right\rangle$, as a structural quantity. The neighboring atoms are determined using the cutoff distance given as 1.4 times the distance corresponding to the first peak in the RDF.

\section{Results and discussion}

\subsection{Si and Ge}\label{3A}

In elemental Si, we perform MD simulations at six temperatures below and above its melting temperature of 1687 K \cite{Melting_point}, i.e., 1600, 1750, 1900, 2050, 2200, and 2350 K, using the Pareto-optimal polynomial MLPs, other interatomic potentials in the literature, and the DFT calculation. In elemental Ge, we also perform MD simulations at six temperatures below and above its melting temperature of 1211 K \cite{Melting_point}, i.e., 1150, 1300, 1450, 1600, 1750, and 1900 K.

\begin{figure}[!tb]
    \centering
    \includegraphics[width=\linewidth]{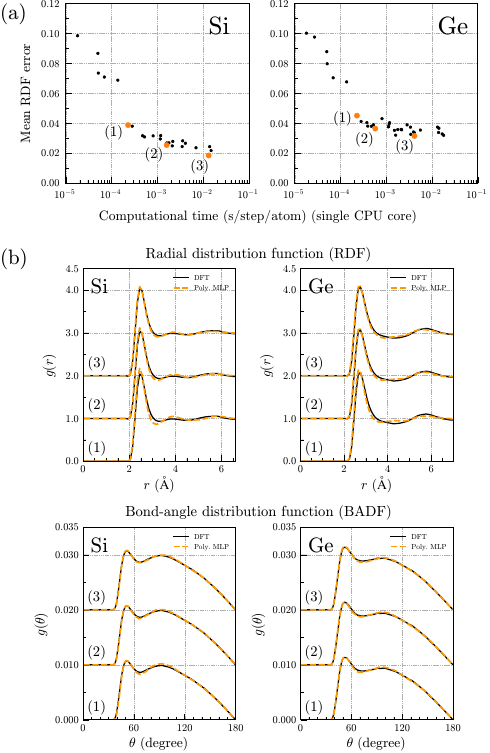}
    \caption{(a) Mean RDF errors of the polynomial MLPs in Si and Ge. The mean RDF error was calculated by averaging the RDF errors at the six temperatures. The computational time on the horizontal axis is regarded as the model complexity of the polynomial MLP. (b) RDFs and BADFs computed using the three polynomial MLPs, (1), (2), and (3), highlighted in (a) at 1750 and 1300 K in Si and Ge, respectively. The RDFs and BADFs at each polynomial MLP are shifted upwards by the amounts of 1.0 and 0.01, respectively. The black solid and orange dotted lines indicate the distribution functions computed using the DFT calculation and the polynomial MLP, respectively.
        }
    \label{fig:sum_smape_time}
\end{figure}

We define the RDF error in a quantitative manner as
\begin{align}
{(\rm{RDF \: error})} = \frac{1}{n_{\rm{bin}}} \sum_{t=1}^{n_{\rm{bin}}} \left| \frac{g_{\rm{pot}}(r_t) - {g_{\rm{DFT}}(r_t)}}{(g_{\rm{pot}}(r_t) + {g_{\rm{DFT}}(r_t)}) /2} \right|,
\end{align}
where $g_{\rm{pot}}(r_t)$ and $g_{\rm{DFT}}(r_t)$ are the frequencies of the single bin centered at $r_t$ in the RDF histogram obtained using an interatomic potential and that obtained using the DFT calculation, respectively. The number of bins is denoted by $n_{\rm{bin}}$. This error metric is known as the symmetric mean absolute percentage error \cite{MAKRIDAKIS1993527}. In this metric, division by zero occurs when both $g_{\rm{pot}}(r_t)$ and $g_{\rm{DFT}}(r_t)$ are equal to zero. Hence, we exclude such bins to calculate the RDF error. Figure \ref{fig:sum_smape_time}(a) shows the mean RDF errors of the polynomial MLPs, calculated by averaging the RDF errors at the six temperatures. As found in Fig. \ref{fig:sum_smape_time}(a), the mean RDF error decreases as the model complexity of polynomial MLP increases, and there is a strong correlation between the mean RDF error and the prediction error for the test dataset.

Figure \ref{fig:sum_smape_time}(b) shows the RDFs and BADFs computed using the three polynomial MLPs highlighted in Fig. \ref{fig:sum_smape_time}(a) at 1750 and 1300 K in Si and Ge, respectively. They are compared with the RDFs and BADFs obtained using the DFT calculation. The RDFs computed using other polynomial MLPs are shown in the supplemental material.  As seen in Fig. \ref{fig:sum_smape_time}(b), the RDFs and BADFs obtained using MLP (1), showing the largest mean RDF error among the three MLPs, are slightly different from those obtained using the DFT calculation. On the other hand, the RDFs and BADFs calculated using the other MLPs almost overlap with those obtained using the DFT calculation.

\begin{figure*}[!p]
    \centering
    \includegraphics[width=0.96\linewidth]{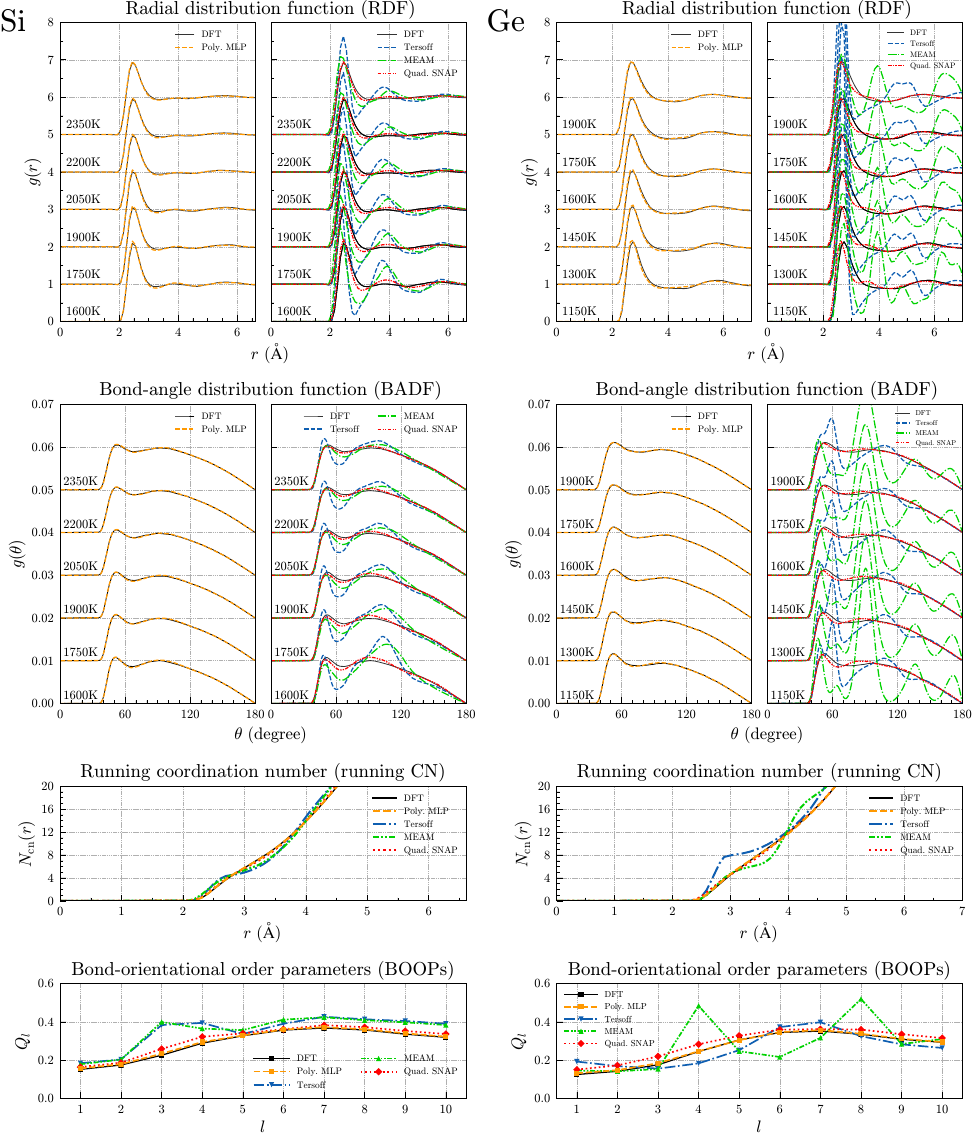}
    \caption{RDFs, BADFs, running CNs, and BOOPs obtained from MD simulations using the polynomial MLPs in elemental Si and Ge. The polynomial MLP showing the lowest mean RDF error is employed for each system. The structural quantities calculated using the DFT calculation, the Tersoff potentials \cite{Si_tersoff, Ge_tersoff}, the MEAM potentials \cite{Si_MEAM, Ge_MEAM}, and the quadratic SNAPs \cite{SNAP} are also shown for comparison. The RDFs and BADFs at each temperature are shifted upwards by the amounts of 1.0 and 0.01, respectively. The running CNs and BOOPs were calculated at 1750 and 1300 K in Si and Ge, respectively, which are close to their melting temperatures. The black solid and orange dotted lines indicate the structural quantities computed using the DFT calculation and the polynomial MLP, respectively. In the legend, Poly. MLP and Quad. SNAP stand for polynomial MLP and quadratic SNAP, respectively.
        }
    \label{fig:sum_SiGe}
\end{figure*}

Figure \ref{fig:sum_SiGe} shows the RDFs and BADFs calculated using the polynomial MLPs at the six temperatures in Si and Ge, as well as those obtained using the DFT calculation. The running CNs and  BOOPs are also calculated at 1750 and 1300 K in Si and Ge, respectively. Here, the polynomial MLP showing the lowest mean RDF error is employed for each system. In Fig. \ref{fig:sum_SiGe}, the structural quantities calculated using the empirical interatomic potentials of the Tersoff potentials \cite{Si_tersoff, Ge_tersoff} and the modified EAM (MEAM) potentials \cite{Si_MEAM, Ge_MEAM} are also shown for comparison. In addition, we calculate the structural quantities using the other MLPs of the quadratic SNAPs \cite{SNAP} that are available in the interatomic potential repository for Si and Ge \cite{OpenKim, NIST} and similar to the polynomial MLPs \cite{doi:10.1063/5.0129045, MachineLearningPotentialRepository}. 

In elemental Si, the structural quantities computed using the polynomial MLP are consistent with the DFT structural quantities at temperatures below and above the melting temperature. The structural quantities calculated using the quadratic SNAP \cite{SNAP} are also close to the DFT structural quantities. However, the predictive power of the quadratic SNAP decreases at temperatures close to the melting temperature. More quantitatively, the RDF errors computed using the polynomial MLP and the quadratic SNAP at the lowest temperature among the six temperatures are 0.021 and 0.105, respectively, while those at the highest temperature among the six temperatures are 0.018 and 0.058, respectively. In temperatures close to the melting temperature, more complex descriptions of the potential energy should be required than those at higher temperatures, where the detailed shape of the potential energy is less important. Regarding the empirical potentials, the structural quantities computed using the Tersoff \cite{Si_tersoff} and MEAM \cite{Si_MEAM} potentials are similar to but inconsistent with the DFT structural quantities. In particular, ghost second peaks are recognized in the RDFs, and the peak intensities of the BADFs differ from those of the DFT calculation. In addition, the running CNs deviate slightly from those of the DFT calculation, and the BOOPs are overestimated for small $l$ values.

In elemental Ge, the structural quantities computed using the polynomial MLP agree with the DFT structural quantities at all six temperatures. The structural quantities calculated using the quadratic SNAP \cite{SNAP} are also close to the DFT structural quantities. However, the predictive power of the quadratic SNAP decreases at temperatures close to the melting temperature, as seen in the case of Si. The RDF errors computed using the polynomial MLP and the quadratic SNAP at the lowest temperature among the six temperatures are 0.051 and 0.125, whereas those at the highest temperature among the six temperatures are 0.033 and 0.059, respectively. Regarding the empirical potentials, the structural quantities computed using the Tersoff \cite{Ge_tersoff} and MEAM \cite{Ge_MEAM} potentials fail to reconstruct the peak positions and intensities of the RDFs and BADFs. The running CNs calculated using the Tersoff and MEAM potentials exhibit plateaus at around 3.0 $\rm{\AA}$ and 3.5 $\rm{\AA}$, respectively, and these $r$ values correspond to the first minimum positions in the RDFs. These plateaus are not found in the running CN of the DFT calculation. The BOOPs calculated using the MEAM potential exhibit two peaks, which differ from the BOOPs obtained using the DFT calculation. The current polynomial MLPs are confirmed to exhibit high predictive power for the liquid structural properties in these systems. On the other hand, the empirical potentials fail to predict the liquid structural properties accurately.

Note that we regard liquid structural quantities obtained from our DFT calculations as the correct ones throughout this study and then compare structural quantities calculated using various interatomic potentials. However, it is challenging to fairly compare the predictive powers of the current polynomial MLPs and interatomic potentials developed elsewhere because different training datasets are regarded as the correct datasets. When an interatomic potential is developed from a DFT training dataset, the interatomic potential depends on the computational method and conditions of the DFT calculation, including the selection of the exchange-correlation functional and PAW potential. Also, some interatomic potentials were developed using experimental training data. Such interatomic potentials can include deviations from our DFT calculations. We have employed typical computational settings for the DFT calculation; hence, it may be fair to say that the present polynomial MLPs can predict liquid structural properties with the same accuracy as those of typical DFT calculations.

\subsection{Li, Be, Na, and Mg}

\begin{figure}[!tb]
    \centering
    \includegraphics[width=\linewidth]{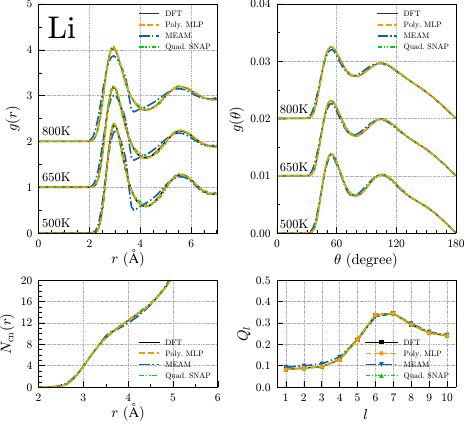}
    \caption{Structural quantities obtained from MD simulations using the polynomial MLP in elemental Li. The polynomial MLP showing the lowest mean RDF error was employed. The structural quantities calculated using the DFT calculation, the MEAM potential \cite{Li_MEAM}, and the quadratic SNAP \cite{SNAP} are also shown for comparison. The RDFs and BADFs shown in the top-left and top-right panels, respectively, were obtained at three temperatures above the melting temperature. The RDFs and BADFs at each temperature are shifted upwards by the amounts of 1.0 and 0.01, respectively. The running CNs and BOOPs shown in the bottom-left and bottom-right panels, respectively, were calculated at the lowest temperature among the three temperatures, which was closest to the melting temperature. The black solid and orange dotted lines indicate the structural quantities computed using the DFT calculation and the polynomial MLP, respectively. In the legend, Poly. MLP and Quad. SNAP stand for polynomial MLP and quadratic SNAP, respectively.}
    \label{fig:sum_Li}
\end{figure}

\begin{figure}[!tb]
    \centering
    \includegraphics[width=\linewidth]{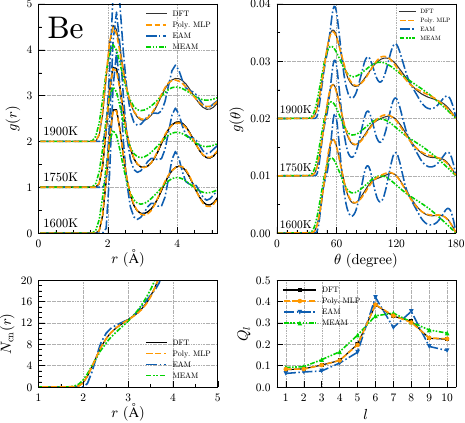}
    \caption{Structural quantities calculated using the DFT calculation, the polynomial MLP, the EAM potential \cite{Be_EAM}, and the MEAM potential \cite{BeO_MEAM} in elemental Be.}
    \label{fig:sum_Be}
\end{figure}

\begin{figure}[!tb]
    \centering
    \includegraphics[width=\linewidth]{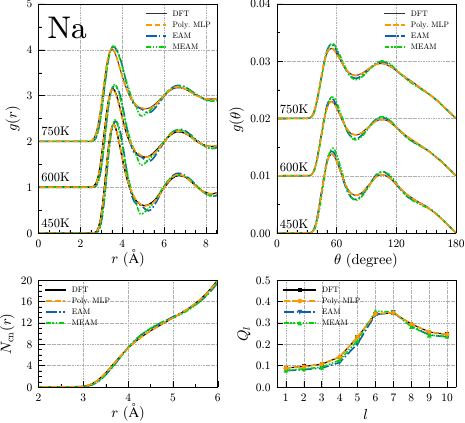}
    \caption{Structural quantities calculated using the DFT calculation, the polynomial MLP, the EAM potential \cite{Na_EAM}, and the MEAM potential \cite{Na_MEAM_Sn_MEAM2} in elemental Na.}
    \label{fig:sum_Na}
\end{figure}

\begin{figure}[!tb]
    \centering
    \includegraphics[width=\linewidth]{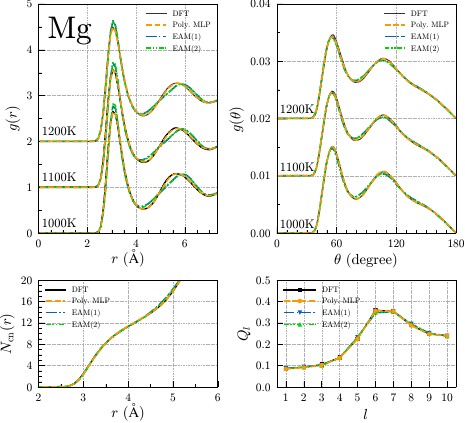}
    \caption{Structural quantities calculated using the DFT calculation, the polynomial MLP, and the EAM potentials \cite{Mg_EAM, Mg_EAM2} in elemental Mg.}
    \label{fig:sum_Mg}
\end{figure}

Figures \ref{fig:sum_Li}, \ref{fig:sum_Be}, \ref{fig:sum_Na}, and \ref{fig:sum_Mg} show the structural quantities calculated from MD simulations using the polynomial MLPs for Li, Be, Na, and Mg, respectively. The RDFs and BADFs are computed at three temperatures above the melting temperatures of 454, 1560, 371, and 923 K \cite{Melting_point} in elemental Li, Be, Na, and Mg, respectively. The running CNs and BOOPs at the lowest temperature among the three temperatures, which is the closest to the melting temperature, are also shown. The structural quantities of the polynomial MLPs are consistent with the DFT structural quantities in Li, Be, Na, and Mg. 

In elemental Li, the RDFs of the MEAM potential \cite{Li_MEAM} slightly deviate from those of the DFT calculation. However, the other structural quantities of the MEAM potential and all structural quantities of the quadratic SNAP \cite{SNAP} are compatible with the DFT structural quantities. In elemental Na, the structural quantities of the EAM \cite{Na_EAM} and MEAM \cite{Na_MEAM_Sn_MEAM2} potentials are close to the DFT structural quantities. In elemental Mg, the EAM potentials \cite{Mg_EAM, Mg_EAM2} and DFT calculation yield consistent structural quantities.

Although the empirical potentials in Li, Na, and Mg accurately predict the liquid structural quantities, the structural quantities computed using the empirical potentials of the EAM \cite{Be_EAM} and MEAM \cite{BeO_MEAM} potentials are inconsistent with the DFT structural quantities in Be. Although the EAM and MEAM potentials qualitatively predict correlations in the RDFs, the peak intensities in the RDFs differ from those of the DFT calculation. Also, three peaks recognized between 45 and 135 degrees in the BADFs of the EAM potential are not found in those of the DFT calculation. The shape of the BADFs calculated using the MEAM potential does not exhibit such ghost peaks, but the peak intensities of the MEAM potential and the DFT calculation are different. In addition, the running CNs and BOOPs computed using the EAM and MEAM potentials differ from those computed using the DFT calculation.

The RDFs computed using all other Pareto-optimal polynomial MLPs at the lowest temperature among the three temperatures are shown in the supplemental material. Most of the polynomial MLPs yield accurate RDFs, except for simplistic polynomial MLPs. Although here we demonstrate only the liquid quantities calculated using the polynomial MLP with the lowest mean RDF error, the above discussion on the predictive power for liquid structures is independent of the selection of the polynomial MLP.

\subsection{Ti, V, and Cr}

\begin{figure}[!tb]
    \centering
    \includegraphics[width=\linewidth]{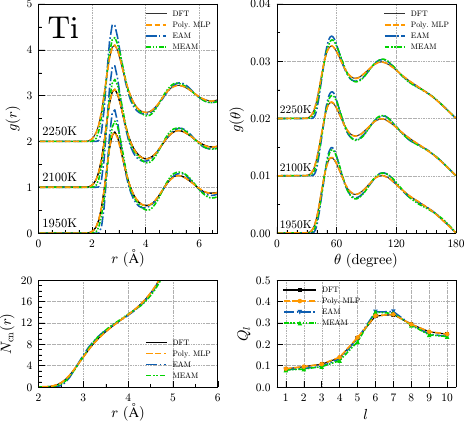}
    \caption{Structural quantities calculated using the DFT calculation, the polynomial MLP, EAM potential \cite{Ag_EAM2_Pb_EAM_Ti_EAM}, and MEAM potential \cite{Ti_MEAM} in elemental Ti.}
    \label{fig:sum_Ti}
\end{figure}

\begin{figure}[!tb]
    \centering
    \includegraphics[width=\linewidth]{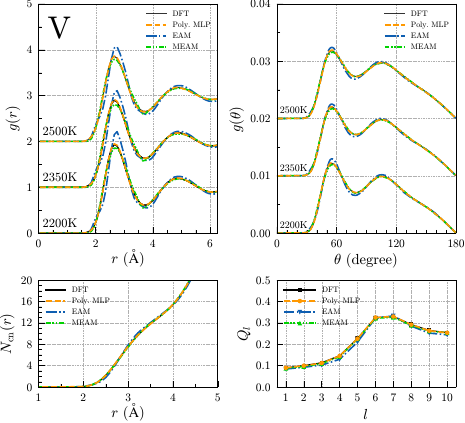}
    \caption{Structural quantities calculated using the DFT calculation, the polynomial MLP, EAM potential \cite{V_EAM}, and MEAM potential \cite{Cr_V_MEAM} in elemental V.}
    \label{fig:sum_V}
\end{figure}

\begin{figure}[!tb]
    \centering
    \includegraphics[width=\linewidth]{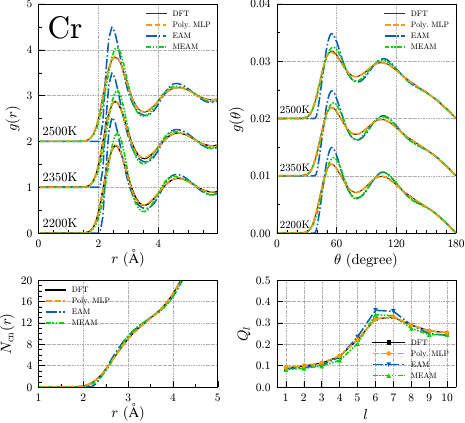}
    \caption{Structural quantities calculated using the DFT calculation, the polynomial MLP, EAM potential \cite{Cr_EAM}, and MEAM potential \cite{Cr_V_MEAM} in elemental Cr.}
    \label{fig:sum_Cr}
\end{figure}

Figures \ref{fig:sum_Ti}, \ref{fig:sum_V}, and \ref{fig:sum_Cr} show the structural quantities calculated from MD simulations at three temperatures using the polynomial MLPs and other empirical potentials for Ti, V, and Cr, respectively. The RDFs and BADFs are computed at three temperatures above the melting temperatures of 1941, 2183, and 2180 K \cite{Melting_point} in elemental Ti, V, and Cr, respectively. The running CNs and BOOPs at the lowest temperature among the three temperatures, which is the closest to the melting temperature, are also shown. The structural quantities of the polynomial MLPs agree with the DFT structural quantities. The RDFs calculated using other polynomial MLPs are shown in the supplemental material and indicate that most of the polynomial MLPs have high predictive power for liquid structural properties.

The EAM \cite{Ag_EAM2_Pb_EAM_Ti_EAM} and MEAM \cite{Ti_MEAM} potentials exhibit similar structural quantities in elemental Ti. However, the peak intensities of the RDFs and BADFs differ slightly from the DFT ones. In elemental V, the RDFs of the EAM \cite{V_EAM} potential slightly deviate from those of the DFT calculation, while the MEAM \cite{Cr_V_MEAM} potential achieves accurate predictions for all structural quantities. The running CN and BOOPs of the MEAM potential \cite{Cr_V_MEAM} are close to the DFT structural quantities in elemental Cr. However, the peak intensities of the RDFs and BADFs differ slightly from those computed from the DFT calculation. The EAM potential \cite{Cr_EAM} shows less accurate structural quantities than the MEAM potential.

\subsection{Cu, Ag, and Au}

\begin{figure}[!tb]
    \centering
    \includegraphics[width=\linewidth]{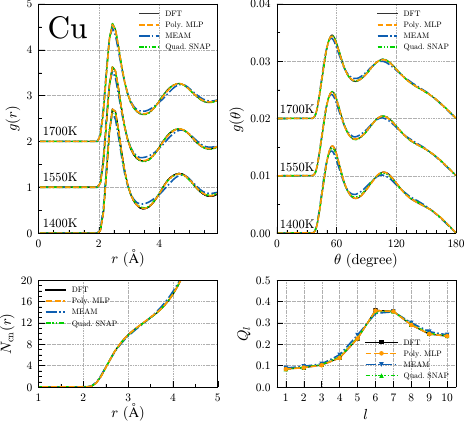}
    \caption{Structural quantities calculated using the DFT calculation, the polynomial MLP, the MEAM potential \cite{Cu_MEAM}, and quadratic SNAP \cite{SNAP} in elemental Cu.}
    \label{fig:sum_Cu}
\end{figure}

\begin{figure}[!tb]
    \centering
    \includegraphics[width=\linewidth]{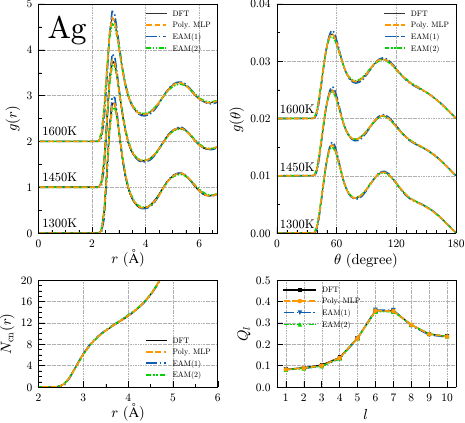}
    \caption{Structural quantities calculated using the DFT calculation, the polynomial MLP, and the EAM potentials \cite{Ag_EAM, Ag_EAM2_Pb_EAM_Ti_EAM} in elemental Ag.}
    \label{fig:sum_Ag}
\end{figure}

\begin{figure}[!tb]
    \centering
    \includegraphics[width=\linewidth]{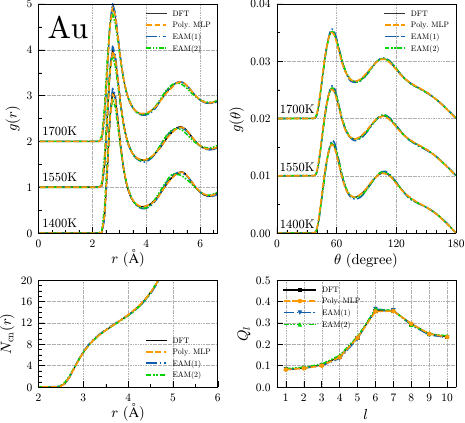}
    \caption{Structural quantities calculated using the DFT calculation, the polynomial MLP, and the EAM potentials \cite{Al_EAM_Au_EAM, Au_EAM2} in elemental Au.}
    \label{fig:sum_Au}
\end{figure}

Figures \ref{fig:sum_Cu}, \ref{fig:sum_Ag}, and \ref{fig:sum_Au} show the structural quantities calculated from MD simulations at three temperatures using the polynomial MLPs and other interatomic potentials for Cu, Ag, and Au, respectively. The RDFs and BADFs are computed at three temperatures above the melting temperatures of 1358, 1235, and 1337 K \cite{Melting_point} in elemental Cu, Ag, and Au, respectively. The running CNs and BOOPs at the lowest temperature among the three temperatures, which is the closest to the melting temperature, are also shown. The structural quantities of the polynomial MLPs are comparable to the DFT structural quantities in Cu, Ag, and Au. The RDFs calculated using other polynomial MLPs are shown in the supplemental material and indicate that most of the polynomial MLPs have high predictive power for liquid structural properties.

In elemental Cu, the structural quantities of the MEAM potential \cite{Cu_MEAM} agree with the DFT ones. The quadratic SNAP \cite{SNAP} also reconstructs the DFT structural quantities accurately. In elemental Ag and Au, all the empirical potentials \cite{Ag_EAM, Ag_EAM2_Pb_EAM_Ti_EAM, Al_EAM_Au_EAM, Au_EAM2} yield accurate structural quantities. 

\subsection{Zn, Cd, and Hg}

\begin{figure}[!tb]
    \centering
    \includegraphics[width=\linewidth]{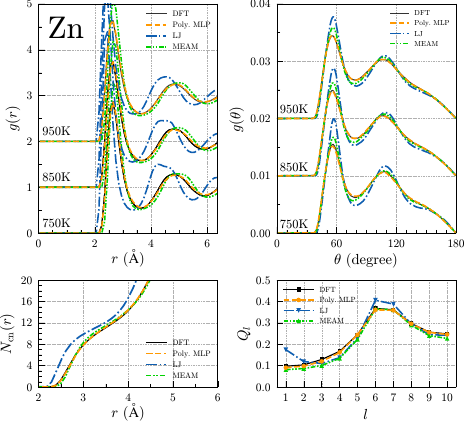}
    \caption{Structural quantities calculated using the DFT calculation, the polynomial MLP, the LJ potential \cite{OpenKIM-MO:959249795837:003}, and MEAM potential \cite{Zn_MEAM} in elemental Zn.}
    \label{fig:sum_Zn}
\end{figure}

\begin{figure}[!tb]
    \centering
    \includegraphics[width=\linewidth]{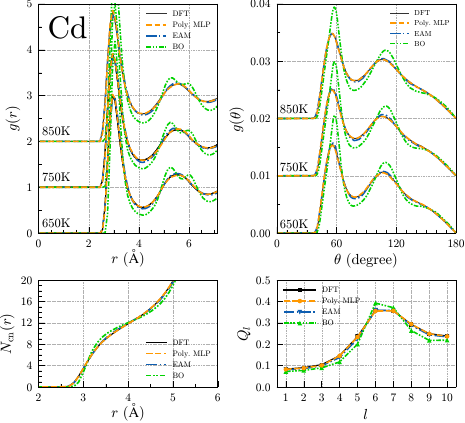}
    \caption{Structural quantities calculated using the DFT calculation, the polynomial MLP, the EAM potential \cite{Cd_EAM}, and BO potential \cite{Cd_BOP} in elemental Cd.}
    \label{fig:sum_Cd}
\end{figure}

\begin{figure}[!tb]
    \centering
    \includegraphics[width=\linewidth]{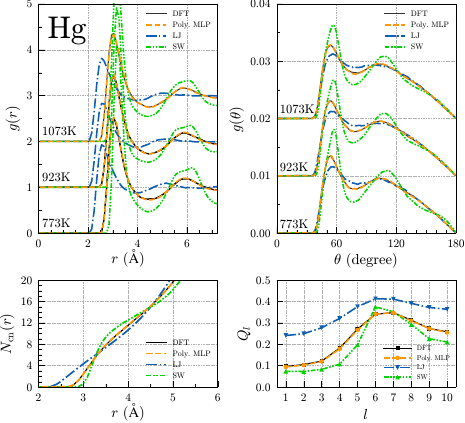}
    \caption{Structural quantities calculated using the DFT calculation, the polynomial MLP, the LJ potential \cite{OpenKIM-MO:959249795837:003}, and SW potential \cite{Hg_SW} in elemental Hg.}
    \label{fig:sum_Hg}
\end{figure}

Figures \ref{fig:sum_Zn}, \ref{fig:sum_Cd}, and \ref{fig:sum_Hg} show the structural quantities calculated from MD simulations at three temperatures using the polynomial MLPs and other empirical potentials for Zn, Cd, and Hg, respectively. The RDFs and BADFs are computed at three temperatures above the melting temperatures of 693, 594, and 234 K \cite{Melting_point} in elemental Zn, Cd, and Hg, respectively. The running CN and BOOPs at the lowest temperature among the three temperatures, which is the closest to the melting temperature, are also shown. The structural quantities of the polynomial MLPs are consistent with the DFT structural quantities. The RDFs calculated using other polynomial MLPs are shown in the supplemental material and indicate that most of the polynomial MLPs have high predictive power for liquid structural properties.

In elemental Zn, the LJ potential \cite{OpenKIM-MO:959249795837:003} is a simple model with potential energy depending solely on the distance between two atoms; hence, it fails to reconstruct the peak positions of the RDFs, the peak intensities of the BADFs, and the running CN of the DFT calculation. The MEAM potential \cite{Zn_MEAM} shows structural quantities better than the LJ potential but fails to reconstruct the peak positions of the RDFs and the peak intensities of the BADFs, similarly to the LJ potential. In elemental Cd, the BO potential \cite{Cd_BOP} fails to accurately predict the peak positions of the RDFs and the peak intensities of the BADFs. On the other hand, structural quantities computed using the EAM potential \cite{Cd_EAM} agree well with the DFT structural quantities. In elemental Hg, the accuracy of the LJ potential \cite{OpenKIM-MO:959249795837:003} for the structural quantities is worse than in Zn, and in particular, the LJ potential fails to reconstruct the BOOPs. The SW potential \cite{Hg_SW} also fails to predict the DFT structural quantities.

\subsection{Al, Ga, In, and Tl}

\begin{figure}[!tb]
    \centering
    \includegraphics[width=\linewidth]{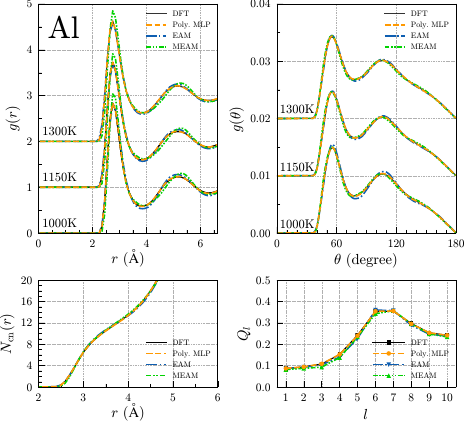}
    \caption{Structural quantities calculated using the DFT calculation, the polynomial MLP, the EAM potential \cite{Al_EAM_Au_EAM}, and MEAM potential \cite{Al_MEAM} in elemental Al.}
    \label{fig:sum_Al}
\end{figure}

\begin{figure}[!tb]
    \centering
    \includegraphics[width=\linewidth]{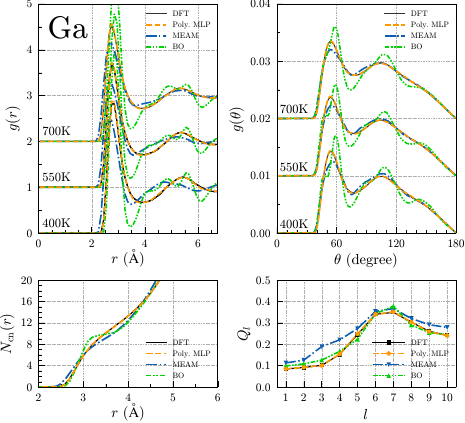}
    \caption{Structural quantities calculated using the DFT calculation, the polynomial MLP, the MEAM potential \cite{GaIn_MEAM}, and BO potential \cite{Ga_BOP} in elemental Ga.}
    \label{fig:sum_Ga}
\end{figure}

\begin{figure}[!tb]
    \centering
    \includegraphics[width=\linewidth]{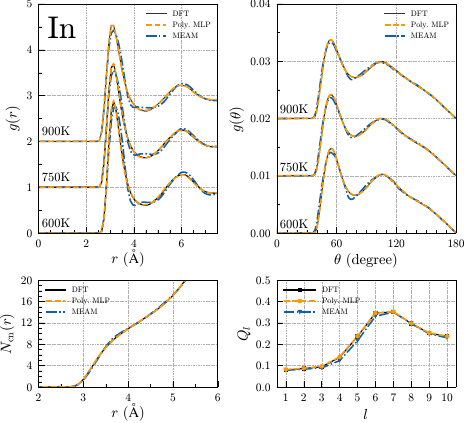}
    \caption{Structural quantities calculated using the DFT calculation, the polynomial MLP, and MEAM potential \cite{In_MEAM} in elemental In.}
    \label{fig:sum_In}
\end{figure}

\begin{figure}[!tb]
    \centering
    \includegraphics[width=\linewidth]{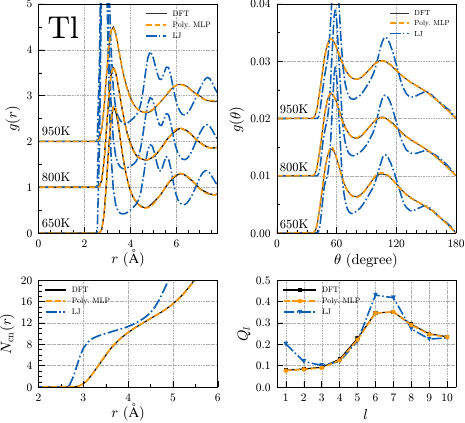}
    \caption{Structural quantities calculated using the DFT calculation, the polynomial MLP, and the LJ potential \cite{OpenKIM-MO:959249795837:003} in elemental Tl.}
    \label{fig:sum_Tl}
\end{figure}

Figures \ref{fig:sum_Al}, \ref{fig:sum_Ga}, \ref{fig:sum_In}, and \ref{fig:sum_Tl} show the structural quantities calculated from MD simulations at three temperatures using the polynomial MLPs and other empirical potentials for Al, Ga, In, and Tl, respectively. The RDFs and BADFs are computed at three temperatures above the melting temperatures of 933, 303, 430, and 577 K \cite{Melting_point} in elemental Al, Ga, In, and Tl, respectively. The running CNs and BOOPs at the lowest temperature among the three temperatures, which is the closest to the melting temperature, are also shown. The structural quantities of the polynomial MLPs agree with the DFT structural quantities. The RDFs calculated using other polynomial MLPs are shown in the supplemental material and indicate that most of the polynomial MLPs have high predictive power for liquid structural properties.

In elemental Al, although the RDFs of the MEAM potential \cite{Al_MEAM} slightly differ from those of the DFT calculation, the other structural quantities of the MEAM potential and all structural quantities of the EAM potential \cite{Al_EAM_Au_EAM} are consistent with the DFT structural quantities. In elemental Ga, the MEAM potential \cite{GaIn_MEAM} cannot accurately reproduce all structure quantities calculated using the DFT calculation. The RDFs, BADFs, and running CNs computed using the BO potential \cite{Ga_BOP} are less accurate than those computed using the MEAM potential. In elemental In, the MEAM potential \cite{In_MEAM} results are close to the DFT structural quantities. In elemental Tl, the LJ potential \cite{OpenKIM-MO:959249795837:003} shows incorrect structural quantities.

\begin{figure}[!tb]
    \centering
    \includegraphics[width=\linewidth]{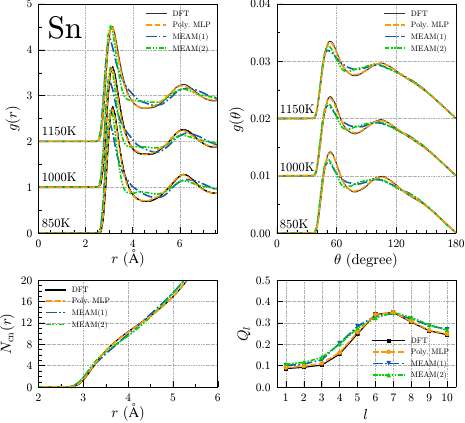}
    \caption{Structural quantities calculated using the DFT calculation, the polynomial MLP, and MEAM potentials \cite{Sn_MEAM, Na_MEAM_Sn_MEAM2} in elemental Sn.}
    \label{fig:sum_Sn}
\end{figure}

\begin{figure}[!tb]
    \centering
    \includegraphics[width=\linewidth]{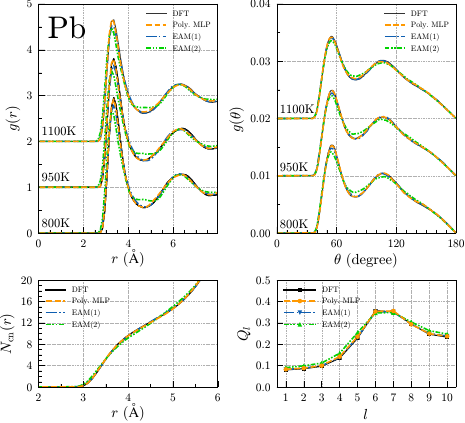}
    \caption{Structural quantities calculated using the DFT calculation, the polynomial MLP, and EAM potentials \cite{Ag_EAM2_Pb_EAM_Ti_EAM, Pb_EAM2} in elemental Pb.}
    \label{fig:sum_Pb}
\end{figure}

\begin{figure}[!tb]
    \centering
    \includegraphics[width=\linewidth]{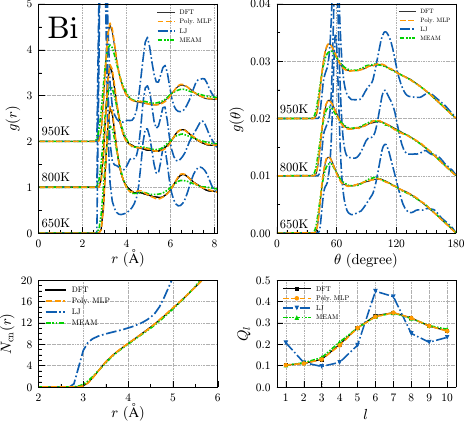}
    \caption{Structural quantities calculated using the DFT calculation, the polynomial MLP, LJ potential \cite{OpenKIM-MO:959249795837:003}, and MEAM potential \cite{Bi_MEAM} in elemental Bi.}
    \label{fig:sum_Bi}
\end{figure}

\begin{table*}[tb]
    \centering
    \caption{RDF errors computed using interatomic potentials at the lowest temperature among our temperature settings. Poly. MLP and Quad. SNAP stand for polynomial MLP and quadratic SNAP, respectively.
      }
    \label{tab:RDF_error}
    \begin{ruledtabular}
    \begin{tabular}{llc|llc|llc}
        Element & Potential & RDF error &  Element & Potential & RDF error &  Element & Potential & RDF error \\
        \hline
        Li & Poly. MLP & 0.007 & Cr & Poly. MLP & 0.011 & Sn & Poly. MLP & 0.035 \\
        & Quad. SNAP \cite{SNAP}  & 0.036 &  & MEAM \cite{Cr_V_MEAM} & 0.141 &  & MEAM(1) \cite{Sn_MEAM} & 0.175 \\
        & MEAM \cite{Li_MEAM}  & 0.183 &  & EAM \cite{Cr_EAM} & 0.380 &  & MEAM(2) \cite{Na_MEAM_Sn_MEAM2} & 0.189 \\
        \hline
        Be & Poly. MLP  & 0.030 & Cu & Poly. MLP & 0.014 & Au & Poly. MLP & 0.014 \\
        & MEAM \cite{BeO_MEAM}  & 0.339 & & Quad. SNAP \cite{SNAP} & 0.033 &  & EAM(1) \cite{Al_EAM_Au_EAM} & 0.041 \\
        & EAM \cite{Be_EAM}  & 0.418 &  & MEAM \cite{Cu_MEAM} & 0.097 &  & EAM(2) \cite{Au_EAM2} & 0.064 \\
        \hline
        Na & Poly. MLP  & 0.022 & Zn & Poly. MLP & 0.041 & Hg & Poly. MLP & 0.018 \\
        & MEAM \cite{Na_MEAM_Sn_MEAM2} & 0.135 &  & MEAM \cite{Zn_MEAM} & 0.210 &  & SW \cite{Hg_SW} & 0.402 \\
        & EAM \cite{Na_EAM}  & 0.147 &  & LJ \cite{OpenKIM-MO:959249795837:003} & 0.431 &  & LJ \cite{OpenKIM-MO:959249795837:003} & 0.438 \\
        \hline
        Mg & Poly. MLP  & 0.017 & Ga & Poly. MLP & 0.019 & Tl & Poly. MLP & 0.019 \\
        & EAM(2) \cite{Mg_EAM2} & 0.128 &  & MEAM \cite{GaIn_MEAM} & 0.288 &  & LJ \cite{OpenKIM-MO:959249795837:003} & 0.685 \\
        & EAM(1) \cite{Mg_EAM} & 0.130 &  & BO \cite{Ga_BOP} & 0.361 &  &  &  \\
        \hline
        Al & Poly. MLP  & 0.008 & Ge & Poly. MLP & 0.051 & Pb & Poly. MLP & 0.028 \\
        & EAM \cite{Al_EAM_Au_EAM} & 0.058 &  & Quad. SNAP \cite{SNAP} & 0.125 &  & EAM(1) \cite{Ag_EAM2_Pb_EAM_Ti_EAM} & 0.089 \\
        & MEAM \cite{Al_MEAM} & 0.118 &  & Tersoff \cite{Ge_tersoff} & 0.377 &  & EAM(2) \cite{Pb_EAM2} & 0.176 \\
        & &  &  & MEAM \cite{Ge_MEAM} & 0.491 &  &  &  \\
        \hline
        Si & Poly. MLP  & 0.021 & Ag & Poly. MLP & 0.016 & Bi & Poly. MLP & 0.023 \\
        & Quad. SNAP \cite{SNAP} & 0.105 &  & EAM(1) \cite{Ag_EAM} & 0.040 &  & MEAM \cite{Bi_MEAM} & 0.158 \\
        & Tersoff \cite{Si_tersoff} & 0.305 &  & EAM(2) \cite{Ag_EAM2_Pb_EAM_Ti_EAM} & 0.063 &  & LJ \cite{OpenKIM-MO:959249795837:003} & 0.650 \\
        & MEAM \cite{Si_MEAM} & 0.342 &  &  &  &  &  &  \\
        \hline
        Ti & Poly. MLP  & 0.012 & Cd & Poly. MLP & 0.015 &  &  &  \\
        & MEAM \cite{Ti_MEAM} & 0.221 &  & EAM \cite{Cd_EAM} & 0.037  &  &  &  \\
        & EAM \cite{Ag_EAM2_Pb_EAM_Ti_EAM} & 0.288 &  & BO \cite{Cd_BOP} & 0.320 &  &  &  \\
        \cline{1-6}
        V & Poly. MLP  & 0.014 & In & Poly. MLP & 0.013 &  &  &  \\
        & MEAM \cite{Cr_V_MEAM} & 0.096 &  & MEAM \cite{In_MEAM} & 0.073 &  &  &  \\
        & EAM \cite{V_EAM} & 0.156 &  &  &  &  &  &  \\

    \end{tabular}
    \end{ruledtabular}
\end{table*}

\subsection{Sn, Pb, and Bi}

Figures \ref{fig:sum_Sn}, \ref{fig:sum_Pb}, and \ref{fig:sum_Bi} show the structural quantities calculated from MD simulations at three temperatures using the polynomial MLPs and other empirical potentials for Sn, Pb, and Bi, respectively. The RDFs and BADFs are computed at three temperatures above the melting temperatures of 505, 601, and 544 K \cite{Melting_point} in elemental Sn, Pb, and Bi, respectively. The running CNs and BOOPs at the lowest temperature among the three temperatures, which is the closest to the melting temperature, are also shown. The structural quantities of the polynomial MLPs are consistent with the DFT structural quantities. The RDFs calculated using other polynomial MLPs are shown in the supplemental material and indicate that most of the polynomial MLPs have high predictive power for liquid structural properties.

Two MEAM potentials \cite{Sn_MEAM, Na_MEAM_Sn_MEAM2} exhibit running CNs and BOOPs similar to those of the DFT calculation in elemental Sn. However, the RDFs and BADFs of the MEAM potentials differ slightly from those of the DFT calculation. In elemental Pb, one EAM potential \cite{Ag_EAM2_Pb_EAM_Ti_EAM} can predict structural quantities accurately, while another EAM potential \cite{Pb_EAM2} shows less accurate structural quantities. In elemental Bi, the LJ potential \cite{OpenKIM-MO:959249795837:003} fails to reconstruct all structural quantities of the DFT calculation. On the other hand, the running CN and BOOPs computed using the MEAM potential \cite{Bi_MEAM} almost overlap with those obtained using the DFT calculation. However, the MEAM potential cannot accurately reproduce the RDFs and BADFs calculated using the DFT calculation.

\subsection{RDF errors}

Table \ref{tab:RDF_error} summarizes the RDF errors computed using interatomic potentials at the lowest temperature among our temperature settings. In all the systems, the RDF error for the polynomial MLP is the smallest, ranging approximately from 0.01 to 0.05. In elemental Li, Al, Cu, Ag, Cd, and Au, the values of the RDF errors are less than 0.06 for some interatomic potentials other than the polynomial MLP. These values of the RDF error are comparable to the mean RDF error shown in Fig. \ref{fig:sum_smape_time}(a). As can be seen in Figs. \ref{fig:sum_smape_time}(a) and (b), the interatomic potentials that exhibit RDF errors less than 0.06 reproduce the RDFs obtained from our DFT calculations accurately. In elemental Na, Mg, Si, V, Ge, In, and Pb, some interatomic potentials other than the polynomial MLP exhibit RDF errors ranging from 0.06 to 0.14. The accuracy of these potentials corresponds to that of simplistic polynomial MLPs with low computational costs. The empirical potentials show RDF errors larger than 0.14 in elemental Li, Be, Si, Ti, Cr, Zn, Ga, Ge, Sn, Hg, Tl, and Bi. These potentials fail to predict the RDFs accurately.

As pointed out in Sec. \ref{3A}, interatomic potentials other than the polynomial MLPs sometimes include systematic deviations from our DFT calculations because of the use of different training datasets. However, some empirical potentials yield RDFs that are totally different from RDFs obtained using our DFT calculations with typical computational settings. On the other hand, the polynomial MLPs exhibit minor RDF errors in all the elemental systems, which indicates that the polynomial MLPs can predict liquid structural properties with the same accuracy as those of reasonable DFT calculations.

\begin{table}[!tb]
\centering
\caption{\label{tab:table_PAW_potential} Electronic configurations of selected PAW potentials that are considered as valence electrons.}
\begin{ruledtabular}
\begin{tabular}{cc|cc}
 Element & Valence states & Element & Valence states \\ 
 \hline
Li & $2s^{1}$ & Ga & $4s^{2}4p^{1}$  \\
Be & $2s^{2}$ & Ge & $4s^{2}4p^{2}$  \\
Na & $3s^{1}$ & Ag & $4d^{10}5s^{1}$  \\
Mg & $3s^{2}$ & Cd & $4d^{10}5s^{2}$  \\
Al & $3s^{2}3p^{1}$ & In & $5s^{2}5p^{1}$   \\
Si & $3s^{2}3p^{2}$ & Sn & $5s^{2}5p^{2}$   \\
Ti & $3d^{2}4s^{2}$ & Au & $5d^{10}6s^{1}$  \\
V & $3d^{3}4s^{2}$ & Hg & $5d^{10}6s^{2}$ \\
Cr & $3d^{5}4s^{1}$ & Tl & $6s^{2}6p^{1}$ \\
Cu & $3d^{10}4s^{1}$ & Pb & $6s^{2}6p^{2}$ \\
Zn & $3d^{10}4s^{2}$ & Bi & $6s^{2}6p^{3}$  \\
 \end{tabular}
\end{ruledtabular}
\end{table}

\begin{figure*}[!tb]
    \centering
    \includegraphics[width=\linewidth]{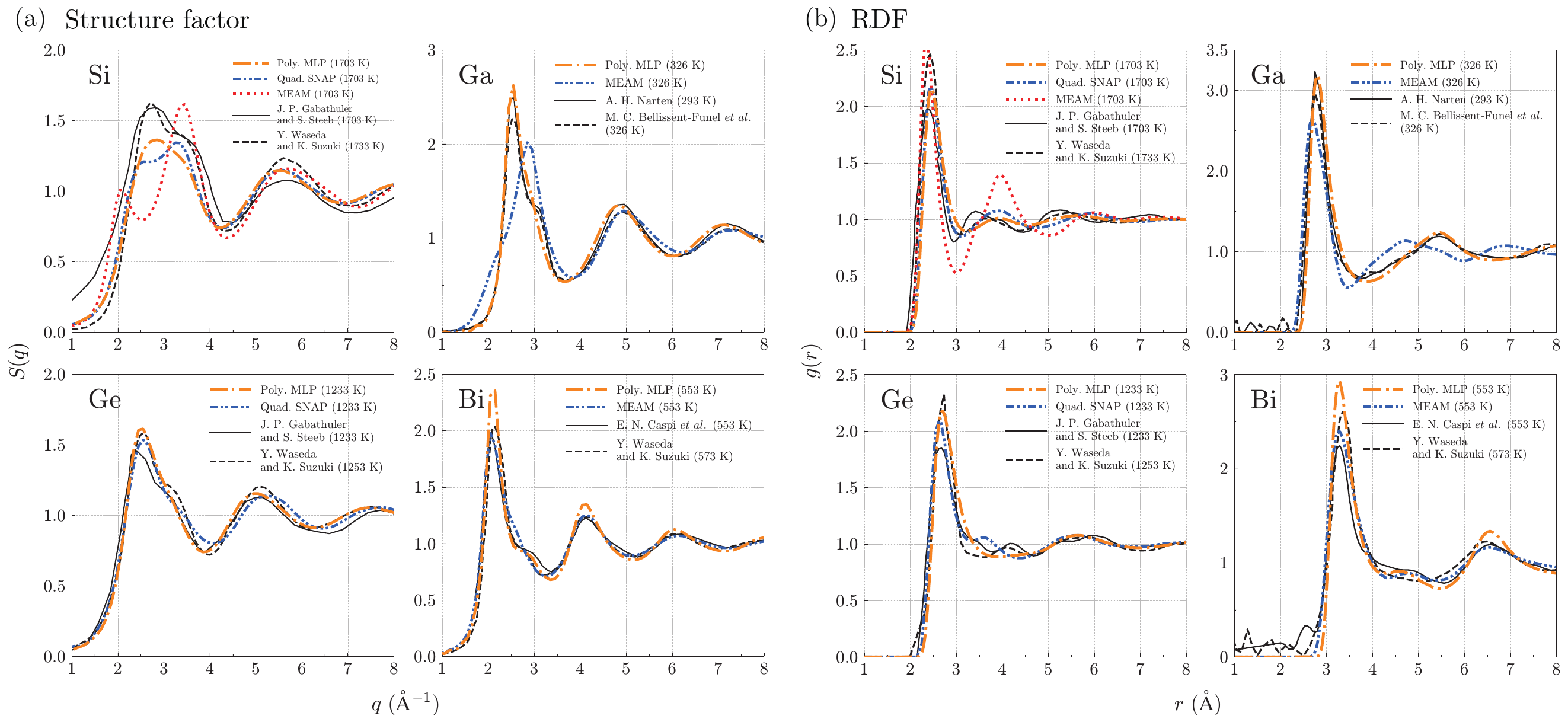}
    \caption{(a) Structure factors computed using interatomic potentials such as the polynomial MLPs in elemental Si, Ga, Ge, and Bi, along with experimental structure factors revealed by X-ray and neutron diffraction investigations \cite{GabathulerSteeb+1979+1314+1319, Waseda1975, 10.1063/1.1677342, PhysRevA.39.6310, Caspi_2012, https://doi.org/10.1002/pssb.2220490132}. (Si \cite{GabathulerSteeb+1979+1314+1319, Waseda1975}, Ga \cite{10.1063/1.1677342, PhysRevA.39.6310}, Ge \cite{GabathulerSteeb+1979+1314+1319, Waseda1975}, Bi \cite{Caspi_2012, https://doi.org/10.1002/pssb.2220490132}).
    (b) RDFs obtained using interatomic potentials in elemental Si, Ga, Ge, and Bi, along with experimental RDFs revealed by X-ray and neutron diffraction investigations \cite{GabathulerSteeb+1979+1314+1319, Waseda1975, 10.1063/1.1677342, PhysRevA.39.6310, Caspi_2012, https://doi.org/10.1002/pssb.2220490132}.}
    \label{fig:experiment_4}
\end{figure*}

\section{Conclusion}

We examined the predictive power of the polynomial MLPs for structural properties in liquid states of 22 elemental systems. Structural quantities such as the RDF and BADF were used to compare the predictive power of the polynomial MLP and other interatomic potentials. The current polynomial MLPs were systematically developed from diverse crystal structures and their derivatives. No structural data in liquid states, such as structural trajectories in MD simulations at high temperatures, were used as training datasets. Nevertheless, they consistently exhibited high predictive power for the liquid structural properties in all 22 elemental systems of diverse chemical bonding properties. On the other hand, empirical potentials failed to predict liquid structural properties in many elemental systems where complex descriptions of the potential energy are required, such as Si, Ge, and Bi. Thus, we can conclude that the polynomial MLPs enable us to efficiently and accurately predict structural and dynamical properties not only in crystalline states but also in liquid and liquid-like disordered states.

\begin{acknowledgments}
This work was supported by a Grant-in-Aid for Scientific Research (A) (Grant Number 21H04621), a Grant-in-Aid for Scientific Research (B) (Grant Number 22H01756), and a Grant-in-Aid for Scientific Research on Innovative Areas (Grant Number 19H05787) from the Japan Society for the Promotion of Science (JSPS).
\end{acknowledgments}

\appendix

\section{Selected PAW potentials}\label{appendixA}

Table \ref{tab:table_PAW_potential} summarizes the electronic configurations of selected PAW potentials that are considered as valence electrons. These PAW potentials were used to construct DFT datasets and perform AIMD simulations in this study.

\section{Comparison with experimental profiles}

\begin{figure*}[!tb]
    \centering
    \includegraphics[width=\linewidth]{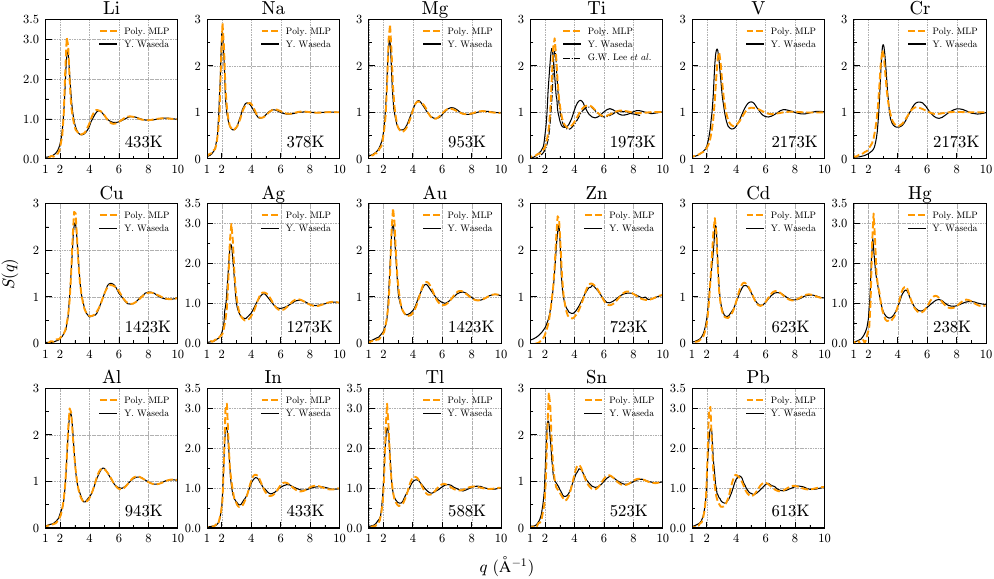}
    \caption{Structure factors computed using the polynomial MLPs along with experimental structure factors revealed by X-ray diffraction investigations  \cite{waseda1980structure, Ti_exp}.}
    \label{fig:other_exp_SQ}
\end{figure*}

\begin{figure*}[!tb]
    \centering
    \includegraphics[width=\linewidth]{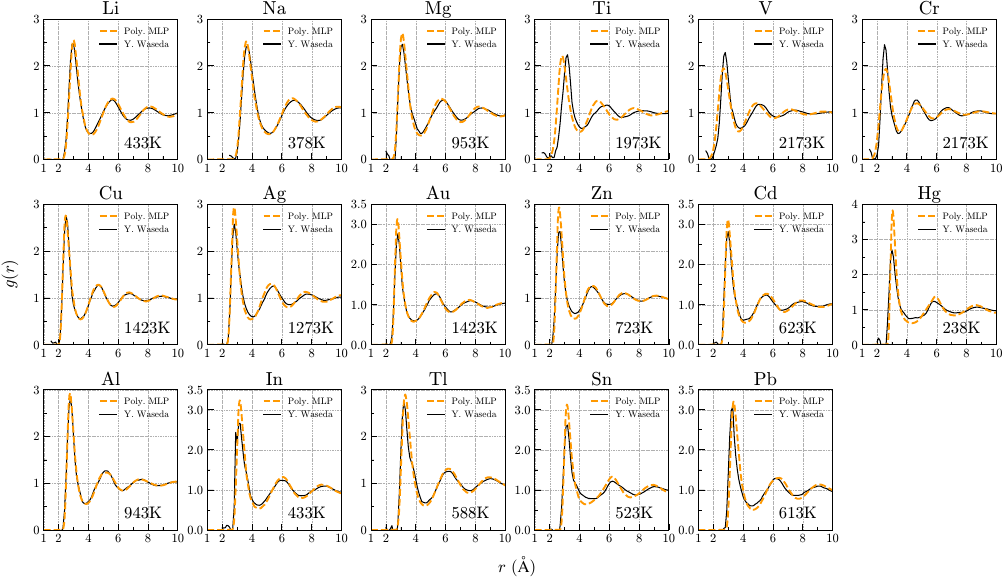}
    \caption{RDFs obtained using polynomial MLPs along with experimental RDFs revealed by X-ray diffraction investigations \cite{waseda1980structure}. }
    \label{fig:other_exp_RDF}
\end{figure*}

\subsection{Si, Ga, Ge, and Bi}\label{appendixB1}

Here, we compare the structure factors and RDFs computed using interatomic potentials such as the polynomial MLP with experimentally reported ones. 
Figure \ref{fig:experiment_4} shows the structure factors and RDFs investigated in X-ray and neutron diffraction experiments in elemental Si, Ga, Ge, and Bi \cite{GabathulerSteeb+1979+1314+1319, Waseda1975, 10.1063/1.1677342, PhysRevA.39.6310, Caspi_2012, https://doi.org/10.1002/pssb.2220490132}. 
Experimental structure factors at similar temperatures are comparable in each system. However, the structure factor profiles show non-negligible differences.
Similarly, experimental RDFs at almost the same temperatures are close to each other, but the first-peak intensities and the positions of the second peaks are inconsistent. 
Although experimental structure factors and RDFs are not uniquely given, we compare the structure factors and RDFs computed using the interatomic potentials with the experimental ones. 
In the supplemental material, the RDFs obtained using the polynomial MLPs are compared with those obtained using DFT calculations, reported in the literature \cite{Be_DFT_RDF, Ti_DFT_RDF, V_DFT_RDF, Cu_DFT_RDF, AgAu_DFT_RDF, ZnCd_DFT_RDF, PhysRevB.55.7539, Al_DFT_RDF, Ga_DFT_RDF1, Ga_DFT_RDF2, kresse_RDF, Si_DFT_RDF2, Pb_DFT_RDF2, Bi_DFT_RDF2}.

Figure \ref{fig:experiment_4} shows the structural factors and the RDFs obtained from MD simulations at 1703, 326, 1233, and 553 K, which are close to the melting temperatures, in elemental Si, Ga, Ge, and Bi, respectively. They were computed using the polynomial MLPs, quadratic SNAPs, and empirical interatomic potentials. In the MD simulations, we employed the simulation cell with 2744 atoms and the computational procedures described in Sec. \ref{2B}. When using the quadratic SNAP \cite{SNAP} for the elemental Ge, we exceptionally employed the simulation cell with 216 atoms because several hundred attempts of MD simulations with larger simulation cells failed. 
The structure factor was obtained by Fourier transforming the corresponding RDF using
\begin{equation}
S(q) = 1 + \frac{4\pi N}{V}  \int_{0}^{L_{\rm{box}}/2} r^2 (g(r) - 1) \frac{{\rm{sin}}(qr)}{qr} {\rm{d}}r ,
\end{equation}
where $L_{\rm{box}}$ represents the simulation box length. This integral was evaluated using a histogram of the RDF with a bin width of 0.1 \AA.

The structure factors of the polynomial MLP and quadratic SNAP \cite{SNAP} in elemental Si are similar to the experimental structure factors. 
However, they do not reconstruct the experimental structure factors around the first peak. 
On the other hand, the structure factors obtained using the polynomial MLP in elemental Ga and Ge are consistent with the experimental structure factors. 
Although the polynomial MLP overestimates the peak intensities in elemental Bi, the structure factor calculated using the polynomial MLP is similar to the experimental one.
The RDFs computed using the polynomial MLP and quadratic SNAP \cite{SNAP} are similar to the experimental RDFs in elemental Si. However, they slightly overestimate the interatomic distances in the experiments. In elemental Ga, the RDF obtained using the polynomial MLP agrees well with the experimental RDFs. In elemental Ge, the first peak position and the RDF at distances larger than 5 $\rm{\AA}$ are consistent with the experimental ones. In contrast, the second peak in the experiments is not clearly observed in the RDF for the polynomial MLP. The second peak is recognized in the RDF for the quadratic SNAP \cite{SNAP}, while its position is different from the experimental ones. In elemental Bi, the RDF computed using the polynomial MLP is similar to but slightly different from the experimental RDFs in terms of peak intensity. Thus, the polynomial MLPs can derive the RDFs close to the experimental RDFs within the range of deviations included in the experimental observations.

\subsection{Other systems}

Figure \ref{fig:other_exp_SQ} shows the structure factor profiles revealed by X-ray diffraction investigations in the other elemental systems at temperatures close to the melting temperatures \cite{waseda1980structure, Ti_exp}. However, no experimental structure factors are available for elemental Be. Figure \ref{fig:other_exp_SQ} also shows the structure factors computed using the polynomial MLPs. The computational procedure is the same as explained in Appendix \ref{appendixB1}. In elemental Li, Na, Cu, Cd, and Al, the structure factor profiles calculated using the polynomial MLPs agree with the experimental ones. In elemental Mg, Ag, Au, Zn, Hg, In, Tl, Sn, and Pb, the first peak intensities of the experimental and computational structure factor profiles are slightly different. However, the polynomial MLPs yield structural factor profiles that are almost the same as the experimental ones. In elemental V and Cr, the structure factors at wave numbers larger than 4 \AA$^{-1}$ slightly differ from the experimental ones. In elemental Ti, the structure factor profile of the polynomial MLP is different from the experimental one investigated by Waseda \cite{waseda1980structure}. However, it is consistent with the experimental one later investigated by Lee \textit{et al.} \cite{Ti_exp}.

Figure \ref{fig:other_exp_RDF} shows the RDFs revealed by X-ray diffraction investigations in the 17 elemental systems at temperatures close to the melting temperatures \cite{waseda1980structure}. 
The RDFs computed using the polynomial MLPs are also shown. The computational procedure is the same as described in Appendix \ref{appendixB1}. 
In elemental Li, Na, Mg, Cu, Au, Cd, Al, and Tl, the RDFs computed using the polynomial MLPs are consistent with the experimental RDFs. 
In elemental Cr, Zn, and In, the RDFs calculated using the polynomial MLPs are similar to the experimental RDFs, while the first peak intensities of the RDFs differ from those of the experimental RDFs. 
The interatomic distances and peak intensities in elemental V, Ag, Hg, Sn, and Pb are slightly different from the experimental ones. 
In elemental Ti, the polynomial MLP fails to reproduce the experimental RDF \cite{waseda1980structure}. However, this experimental RDF was obtained from the structure factor profile that is incompatible with the structure factor profile computed using the polynomial MLP, as shown in Fig. \ref{fig:other_exp_SQ}.
On the other hand, the polynomial MLP has been shown to be capable of reconstructing the other experimental structure factor profile in elemental Ti \cite{Ti_exp}.

\bibliography{references, introduction, liquid-method}

\pagebreak
\widetext
\begin{center}
\textbf{\large Supplemental Material: Predictive power of polynomial machine learning potentials for liquid states in 22 elemental systems}
\end{center}

\setcounter{subsection}{0}
\renewcommand{\thefigure}{S\arabic{figure}}
\renewcommand{\bibnumfmt}[1]{[S#1]}

\subsection{Absolute prediction errors for 86 prototype structures}

\begin{figure*}[!tb]
    \centering
    \includegraphics[width=0.92\linewidth]{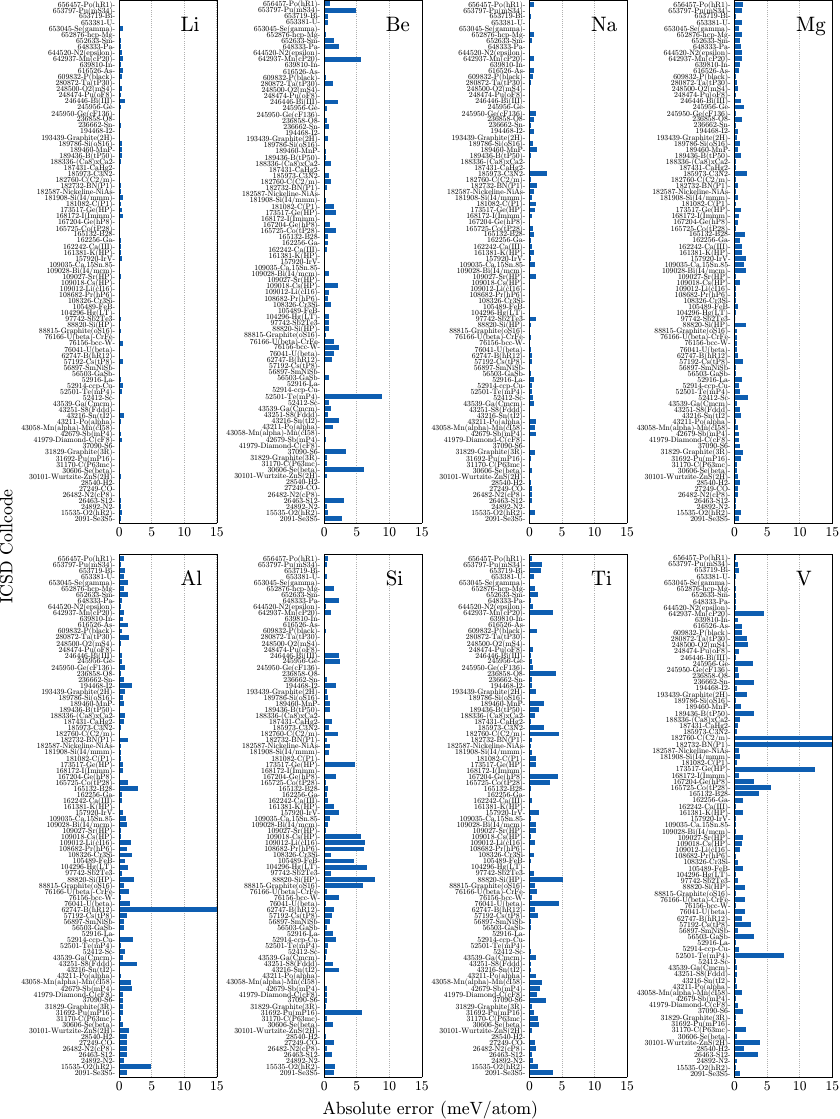}
    \caption{Absolute prediction errors of the cohesive energy for 86 prototype structures in elemental Li, Be, Na, Mg, Al, Si, Ti, and V.
        }
    \label{fig:sum_proto_else1}
\end{figure*}

\begin{figure*}[!tb]
    \centering
    \includegraphics[width=0.92\linewidth]{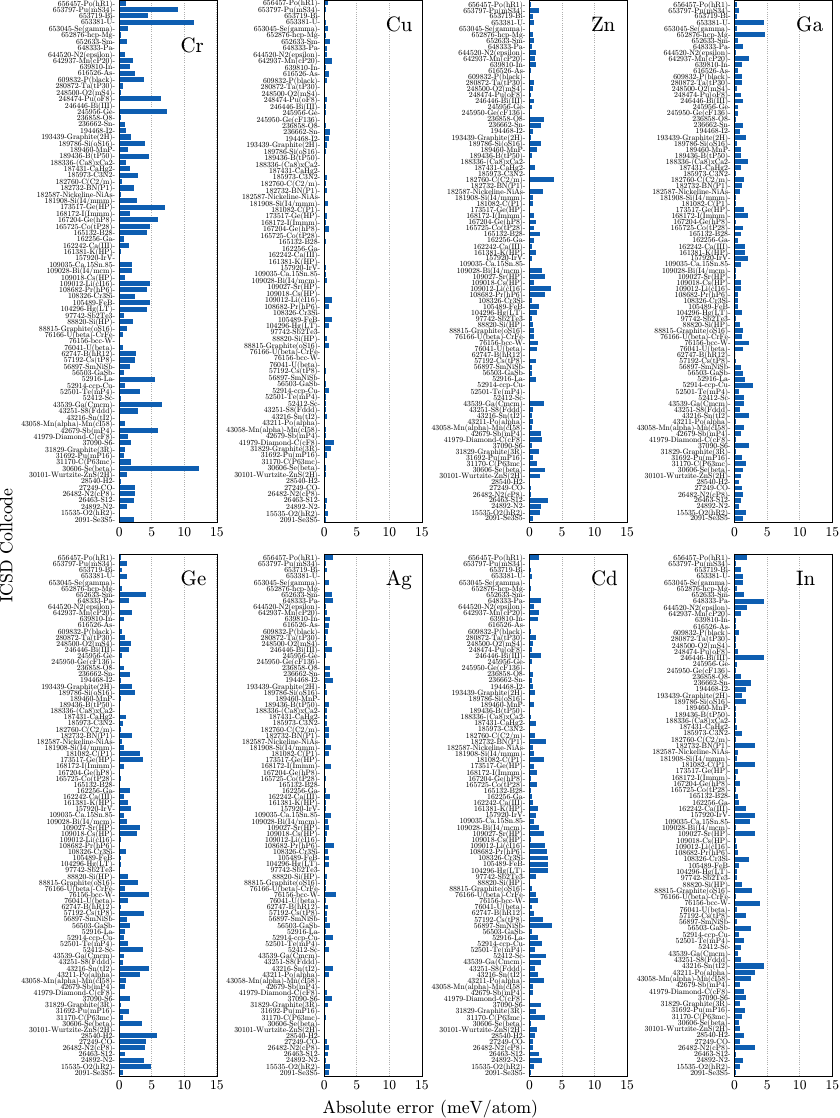}
    \caption{Absolute prediction errors of the cohesive energy for 86 prototype structures in elemental Cr, Cu, Zn, Ga, Ge, Ag, Cd, and In.
        }
    \label{fig:sum_proto_else2}
\end{figure*}

\begin{figure*}[!tb]
    \centering
    \includegraphics[width=0.92\linewidth]{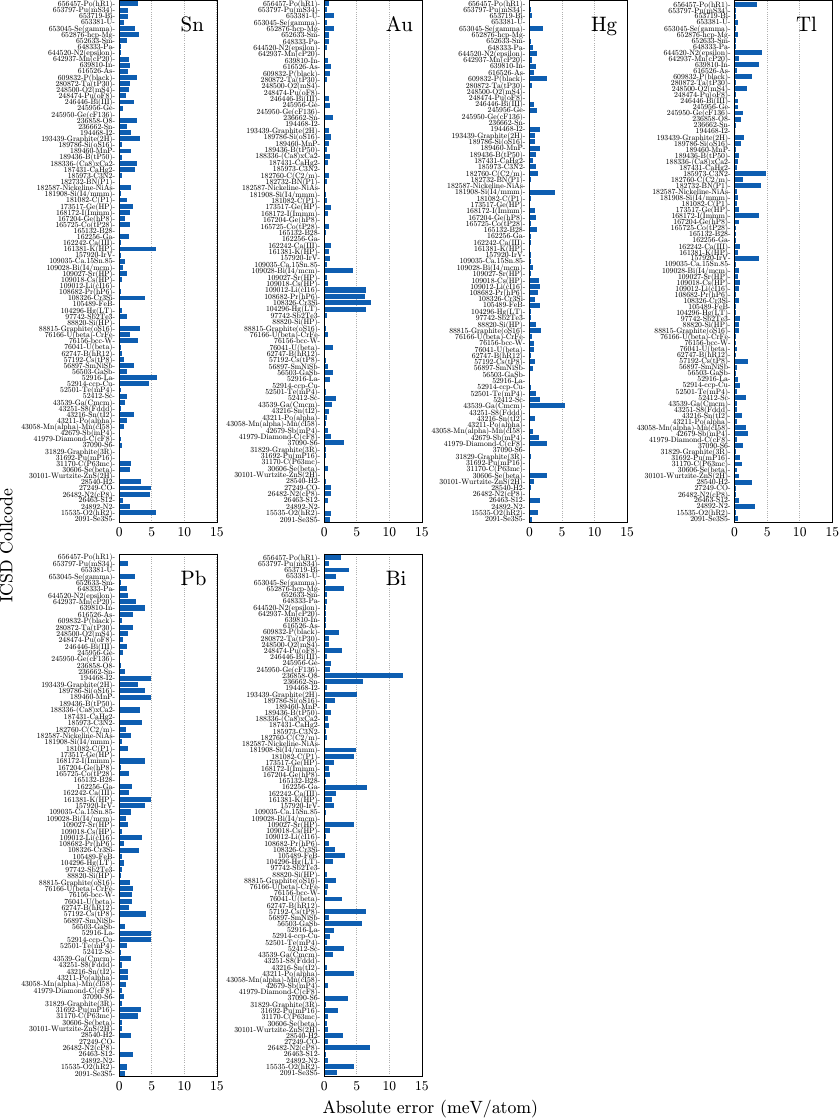}
    \caption{Absolute prediction errors of the cohesive energy for 86 prototype structures in elemental Sn, Au, Hg, Tl, Pb, and Bi.
        }
    \label{fig:sum_proto_else3}
\end{figure*}

Figures \ref{fig:sum_proto_else1}, \ref{fig:sum_proto_else2}, and \ref{fig:sum_proto_else3} show the absolute prediction errors of the cohesive energy for 86 prototype structures in 22 elemental systems, i.e., Li, Be, Na, Mg, Al, Si, Ti, V, Cr, Cu, Zn, Ga, Ge, Ag, Cd, In, Sn, Au, Hg, Tl, Pb, and Bi. 
The polynomial MLP exhibits small errors for almost all prototype structures. 
These results indicate that the polynomial MLP is accurate for many typical structures containing diverse neighborhood environments and coordination numbers.

\subsection{The RDFs calculated using all Pareto-optimal polynomial MLPs}
\begin{figure}[!tb]
    \centering
    \includegraphics[width=0.9\linewidth]{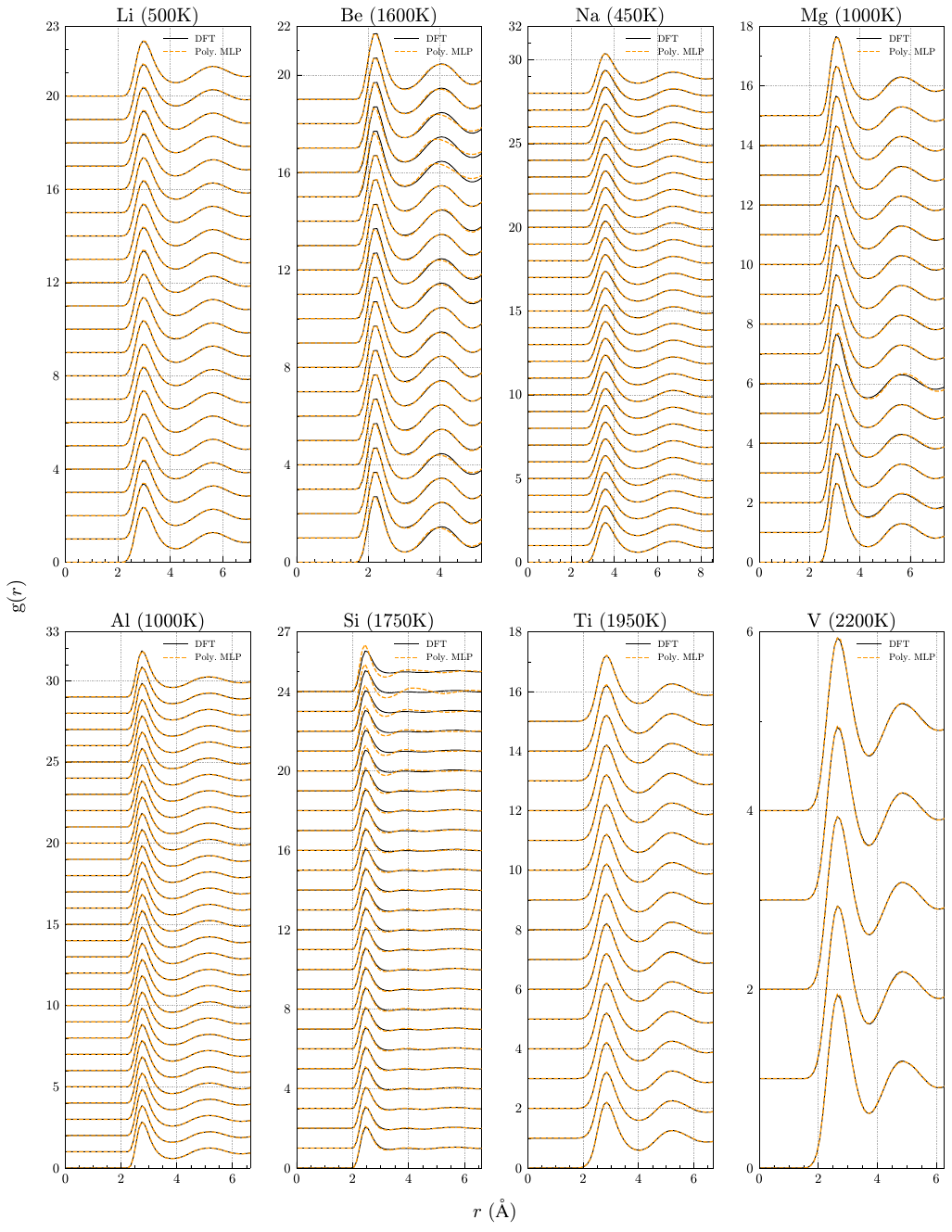}
    \caption{RDFs calculated using all the Pareto-optimal polynomial MLPs at a temperature above and closest to the melting temperature among our temperature settings in elemental Li, Be, Na, Mg, Al, Si, Ti, and V. The RDFs calculated using each polynomial MLP are shifted upwards by the amount of 1.0. The RDFs are shown from top to bottom in the ascending order of the computational cost. In the legend, Poly. MLP stands for polynomial MLP.}
    \label{fig:RDF_sum1}
\end{figure}

\begin{figure}[!tb]
    \centering
    \includegraphics[width=0.9\linewidth]{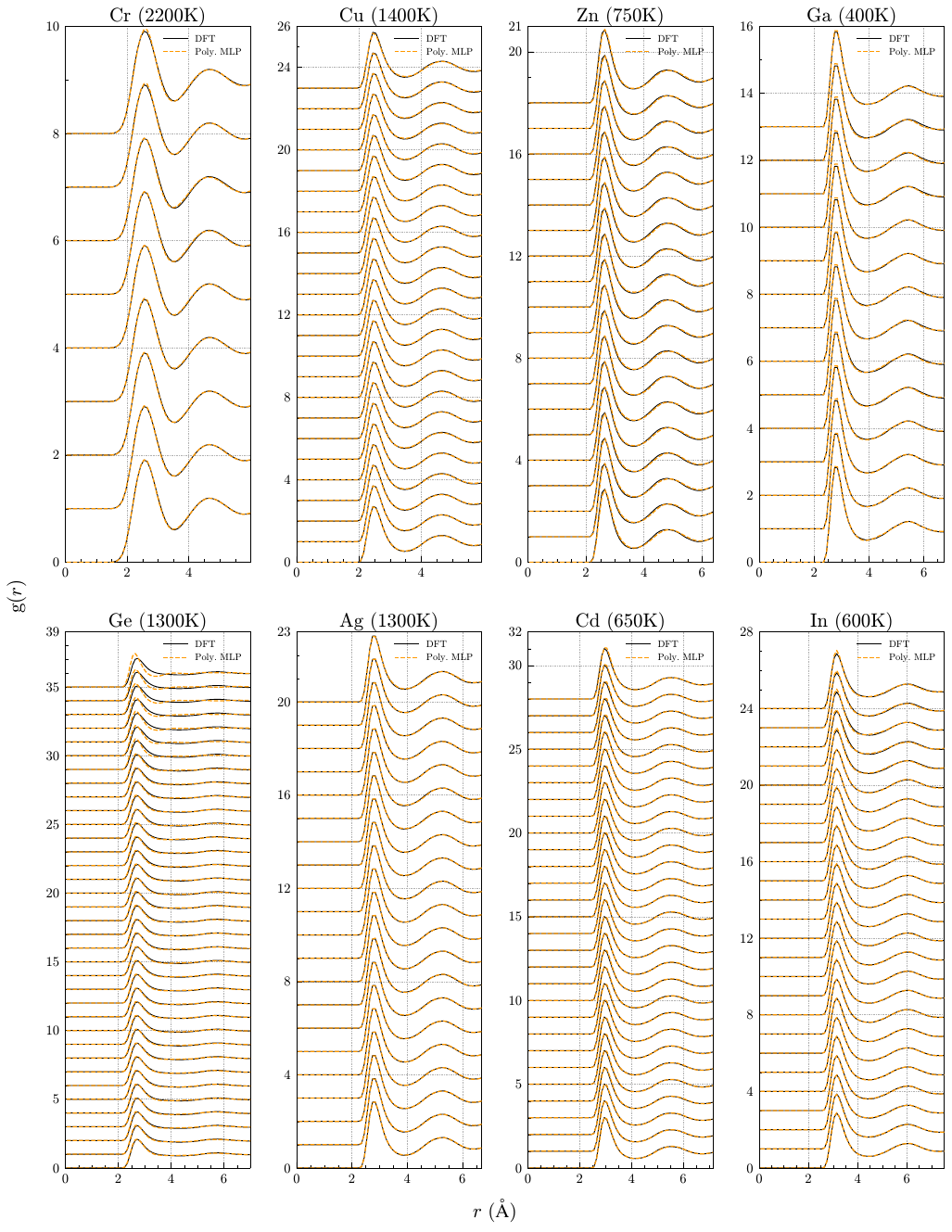}
    \caption{RDFs calculated using all the Pareto-optimal polynomial MLPs at a temperature above and closest to the melting temperature among our temperature settings in elemental Cr, Cu, Zn, Ga, Ge, Ag, Cd, and In. The RDFs calculated using each MLP are shifted upwards by the amount of 1.0. The RDFs are shown from top to bottom in the ascending order of the computational cost. In the legend, Poly. MLP stands for polynomial MLP.}
    \label{fig:RDF_sum2}
\end{figure}

\begin{figure}[!tb]
    \centering
    \includegraphics[width=0.9\linewidth]{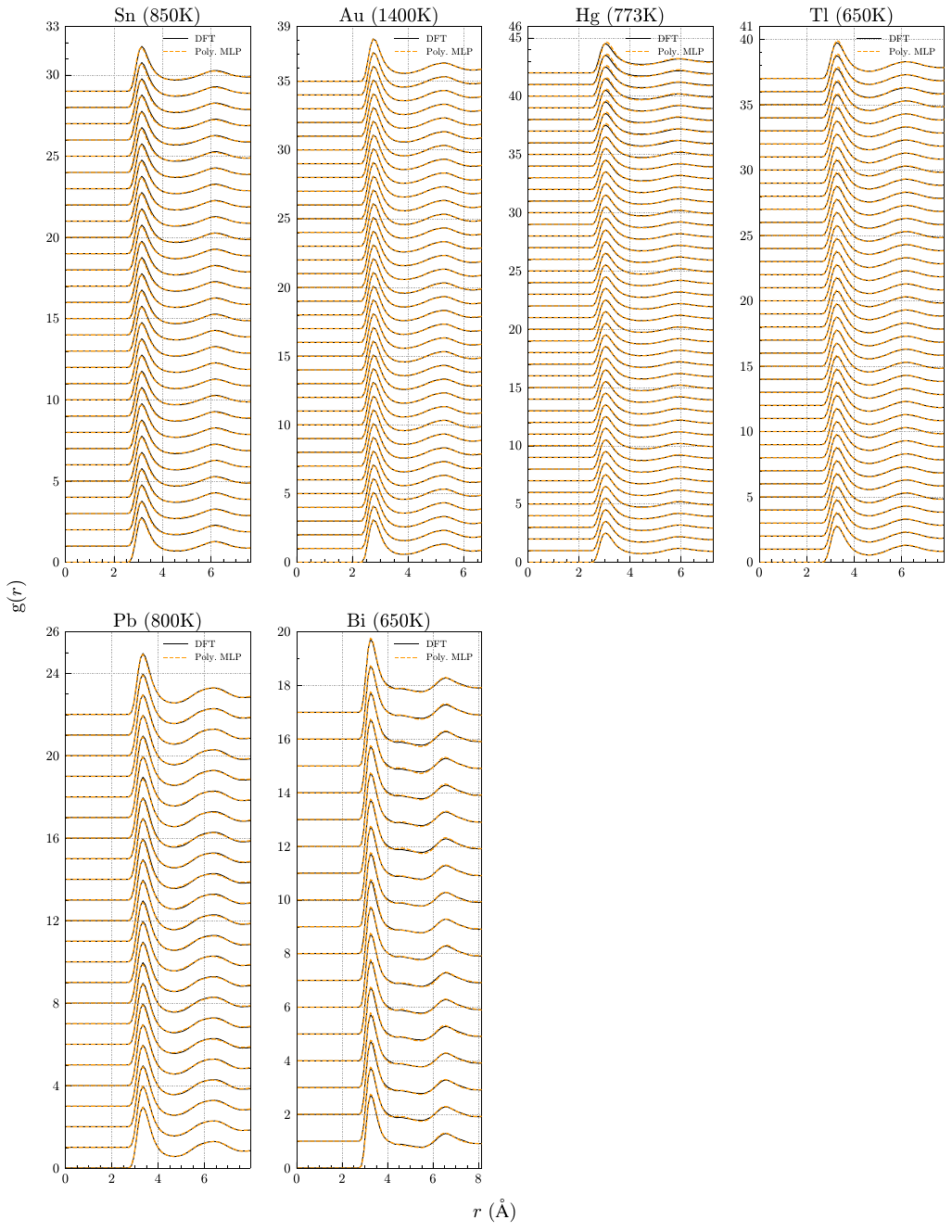}
    \caption{RDFs calculated using all the Pareto-optimal polynomial MLPs at a temperature above and closest to the melting temperature among our temperature settings in elemental Sn, Au, Hg, Tl, Pb, and Bi. The RDFs calculated using each MLP are shifted upwards by the amount of 1.0. The RDFs are shown from top to bottom in the ascending order of the computational cost. In the legend, Poly. MLP stands for polynomial MLP.}
    \label{fig:RDF_sum3}
\end{figure}

Figures \ref{fig:RDF_sum1}, \ref{fig:RDF_sum2}, and \ref{fig:RDF_sum3} exhibit the RDFs calculated using all Pareto-optimal polynomial MLPs at a temperature above and closest to the melting temperature among our temperature settings in elemental Li, Be, Na, Mg, Al, Si, Ti, V, Cr, Cu, Zn, Ga, Ge, Ag, Cd, In, Sn, Au, Hg, Tl, Pb, and Bi. Most of the polynomial MLPs yield accurate RDFs, except for simplistic polynomial MLPs, and the predictive power for liquid structures is independent of the selection of the polynomial MLP.

\subsection{Test DFT calculations using different PAW potentials}

\begin{figure}[!tb]
    \centering
    \includegraphics[width=0.6\linewidth]{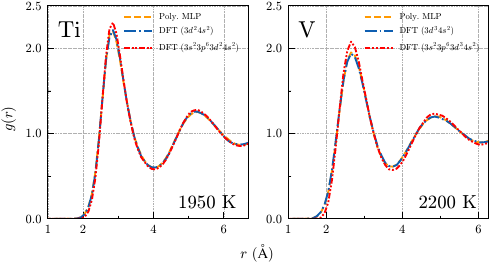}
    \caption{RDFs obtained from DFT calculations with the two kinds of PAW potentials in elemental Ti and V.}
    \label{fig:RDF_semi_TiV}
\end{figure}

We performed AIMD simulations and calculated RDFs using PAW potentials that differ from those applied in the main text for elemental Ti and V. In these PAW potentials, $3s$ and $3p$ semi-core electrons are regarded as valence states. 
We followed the same computational procedures described in Sec. I\hspace{-1.2pt}I B, except that we increased the cutoff energy to 500 eV because these PAW potentials require a higher cutoff energy for accurate DFT calculations.
Figure \ref{fig:RDF_semi_TiV} shows the RDFs obtained from DFT calculations at 1950 and 2200 K in elemental Ti and V, respectively. Although the peak intensities of the RDFs computed using the two kinds of PAW potentials differ slightly in both systems, there are no significant differences.

\begin{figure}[!tb]
    \centering
    \includegraphics[width=\linewidth]{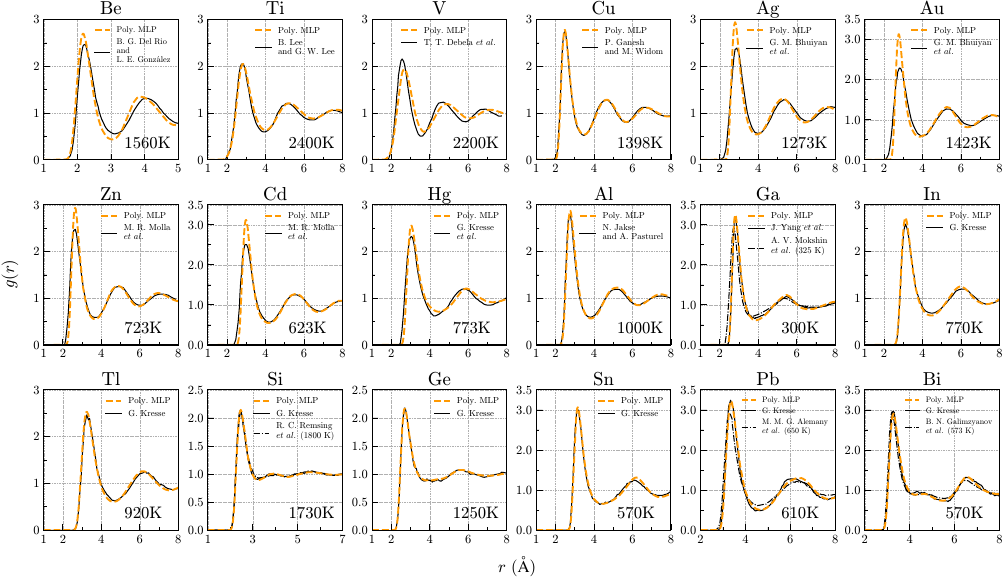}
    \caption{RDFs obtained from the MD simulations using the polynomial MLP in 18 elemental systems, along with RDFs obtained from other DFT calculations \cite{Be_DFT_RDF, Ti_DFT_RDF, V_DFT_RDF, Cu_DFT_RDF, AgAu_DFT_RDF, ZnCd_DFT_RDF, PhysRevB.55.7539, Al_DFT_RDF, Ga_DFT_RDF1, Ga_DFT_RDF2, kresse_RDF, Si_DFT_RDF2, Pb_DFT_RDF2, Bi_DFT_RDF2}. (Be \cite{Be_DFT_RDF}, Ti \cite{Ti_DFT_RDF}, V \cite{V_DFT_RDF}, Cu \cite{Cu_DFT_RDF}, Ag \cite{AgAu_DFT_RDF}, Au \cite{AgAu_DFT_RDF}, Zn \cite{ZnCd_DFT_RDF}, Cd \cite{ZnCd_DFT_RDF}, Hg \cite{PhysRevB.55.7539}, Al \cite{Al_DFT_RDF}, Ga \cite{Ga_DFT_RDF1, Ga_DFT_RDF2}, In \cite{kresse_RDF}, Tl \cite{kresse_RDF}, Si \cite{kresse_RDF, Si_DFT_RDF2}, Ge \cite{kresse_RDF}, Sn \cite{kresse_RDF}, Pb \cite{kresse_RDF, Pb_DFT_RDF2}, Bi \cite{kresse_RDF, Bi_DFT_RDF2}). }
    \label{fig:dft_RDF}
\end{figure}

\subsection{Comparison with RDFs obtained from DFT calculations in the literature}

Here, we compare the RDFs computed using the polynomial MLPs with those obtained from DFT calculations reported in the literature \cite{Be_DFT_RDF, Ti_DFT_RDF, V_DFT_RDF, Cu_DFT_RDF, AgAu_DFT_RDF, ZnCd_DFT_RDF, PhysRevB.55.7539, Al_DFT_RDF, Ga_DFT_RDF1, Ga_DFT_RDF2, kresse_RDF, Si_DFT_RDF2, Pb_DFT_RDF2, Bi_DFT_RDF2}. The RDFs calculated using the polynomial MLPs are obtained through the procedure described in Appendix A. Figure \ref{fig:dft_RDF} shows the comparison between the RDFs obtained from the polynomial MLP and those reported in the literature.

In many systems, the RDFs calculated using the polynomial MLPs are comparable to those calculated using DFT calculations reported in the literature. However, in some systems, there are slight differences in the peak intensities and interatomic distances obtained using the polynomial MLPs as compared to the DFT calculations reported in the literature. Note that the computational procedure and conditions for the DFT calculations in the literature, such as the exchange-correlation functional type, the ensemble type, the number of atoms in the simulation box, and the number of steps, are often different from those used in this study.

\end{document}